\documentclass[12pt]{article}
\usepackage{cite,epsfig,amssymb,amsmath,feynarts}

\usepackage[dvips]{pstcol}

\newenvironment{comment}[1]{}{}
\newcommand{\Eq}[1]{Eq.\,(\ref{#1})}
\newcommand{\Eqs}[1]{Eqs.\,(\ref{#1})}
\newcommand{\Eqsand}[2]{Eqs.\,(\ref{#1}) and (\ref{#2})}

\newcommand{\Fig}[1]{Fig. \ref{#1}}

\newcommand{\beq}{\begin{equation}}                
\newcommand{\eeq}{\end{equation}}        
\newcommand{\bea}{\begin{eqnarray}}               
\newcommand{\eea}{\end{eqnarray}}        
\newcommand{\bdm}{\begin{displaymath}}                 
\newcommand{\edm}{\end{displaymath}}                      
  
\newcommand{\non}{\nonumber}

\newcommand{\smallW}{{\scriptscriptstyle W}}

\newcommand{\ord}{{O}}

\newcommand{\mb}{m_b}

\newcommand{\mw}{M_\smallW}
\newcommand{\MW}{M_\smallW}

\newcommand{\GeV}{\, {\rm GeV}}

\newcommand{\MSbar}{ \overline{\rm MS} }

\newcommand{\gs }{ g }
\newcommand{\as }{ \alpha_s }         
\newcommand{\aem}{ \alpha }   
\newcommand{\f}{\frac}

\newcommand{\Dsl}{D \hspace{-.70em} / \hspace{.25em}}
\newcommand{\dsl}{\partial \hspace{-.60em} / \hspace{.25em}}
\newcommand{\Asl}{A \hspace{-.60em} / \hspace{.25em}}
\newcommand{\Gsl}{G \hspace{-.70em} / \hspace{.25em}}

\newcommand{\Leff}{{\cal L}_{\rm eff}}
\newcommand{\LQCDQED}{{\cal L}_{{\rm QCD} \times {\rm QED}} (u, d, s,
  c, b, e, \mu, \tau)}
\newcommand{\Aeff}{{\cal A}_{\rm eff}}

\newcommand{\SUC}{SU (3)_C}
\newcommand{\UQ}{U (1)_Q}
\newcommand{\GF}{G_F}
\newcommand{\mir}{M}

\newcommand{\dimension}{n}
\newcommand{\ep}{\epsilon}

\newcommand{\btosgamma}{b \to s \gamma}
\newcommand{\btosgluon}{b \to s g}
\newcommand{\btoslpluslminus}{b \to s \ell^+ \ell^-}
\newcommand{\BtoXsgamma}{B \to X_s \gamma}
\newcommand{\BtoXslpluslminus}{B \to X_s \ell^+ \ell^-}
\newcommand{\btosccbar}{b \to s c \bar{c}}
\newcommand{\btos}{b \to s}
\newcommand{\btosgluongluon}{b \to s g g}
\newcommand{\BtoKstargamma}{B \to K^\ast \gamma}
\newcommand{\BtoKstarlpluslminus}{B \to K^\ast \ell^+ \ell^-}
\newcommand{\Btorhogamma}{B \to \rho \gamma}
\newcommand{\Btorholpluslminus}{B \to \rho \ell^+ \ell^-}

\newcommand{\BR}{{\rm BR}}

\newcommand{\eps}{\epsilon}

\newcommand{\zetathree}{\zeta_3}

\newcommand{\EP}{E \hspace{-0.2mm} P}
\newcommand{\EN}{E \hspace{-0.2mm} N}
\newcommand{\PP}{P \hspace{-0.3mm} P}
\newcommand{\PN}{P \hspace{-0.4mm} N}
\newcommand{\PE}{P \hspace{-0.4mm} E}
\newcommand{\NN}{N \hspace{-0.4mm} N}
\newcommand{\EE}{E \hspace{-0.2mm} E}
\newcommand{\NP}{N \hspace{-0.4mm} P}
\newcommand{\NE}{N \hspace{-0.4mm} E}

\newcommand{\ca}{C_A}
\newcommand{\cf}{C_F}
\newcommand{\nf}{N_f}

\newcommand{\AD}{Anomalous Dimension}
\newcommand{\ad}{anomalous dimension}
\newcommand{\ads}{anomalous dimensions}

\newcommand{\etal}{{\it et al}.}

\begin{document}

\allowdisplaybreaks


\thispagestyle{empty}
\rightline{CERN-TH/2003-120}
\rightline{TUM-HEP-515/03}
\rightline{FERMILAB-Pub-03/151-T}
\rightline{\today}
\vspace*{1.2truecm}
\bigskip

\centerline{\LARGE \bf \AD \ Matrix for}
\vspace{.4cm}
\centerline{\LARGE \bf Radiative and Rare Semileptonic}
\vspace{.2cm}
\centerline{\LARGE \bf \boldmath$ B$\unboldmath \ Decays up to Three Loops} 

\vskip1truecm
\centerline{\large\bf Paolo Gambino$^a$, Martin Gorbahn$^b$ and Ulrich
Haisch$^{b,c}$} 
\bigskip
\begin{center}{
{\em $^a$ Theory Division, CERN, \\ CH-1211 Geneve 23, Switzerland }
\\ 
\vspace{.3cm}
{\em $^b$ Physik Department, Technische Universit\"at M\"unchen, \\  
D-85748 Garching, Germany} \\
\vspace{.3cm}
{\em $^c$ Theoretical Physics Department, Fermilab, \\ Batavia, IL
60510, USA}    
}\end{center}
\vspace{1.5cm}

\centerline{\bf Abstract}
\vspace{1.cm}
We compute the complete $\ord (\as^2)$ \ad \ matrix relevant for the
$\btosgamma$, $\btosgluon$ and $\btoslpluslminus$ transitions in the
standard model and some of its extensions. For radiative decays we
confirm the results of Misiak and M\"unz, and of 
Chetyrkin, Misiak and M\"unz. The $\ord (\as^2)$ mixing of four-quark
into  semileptonic operators is instead a new result and represents
one of the last missing ingredients of the
next-to-next-to-leading-order analysis of rare semileptonic $B$ meson
decays.      
\vspace*{2.0cm}

\newpage
\section{Introduction}

Inclusive radiative $B$ decays place very important constraints on
models of physics beyond the Standard Model (SM). The present 
experimental world average for the branching fraction of the radiative
$\BtoXsgamma$   
decay is \cite{exp} 
\beq
\BR ( \BtoXsgamma )_{\rm exp} = \left ( 3.34 \pm 0.38 \right ) \times
10^{-4} \, ,
\eeq
while the most recent SM prediction is
\cite{gambino-misiak,misiak-buras}  
\beq
\BR ( \BtoXsgamma )_{\rm th} = \left ( 3.70 \pm 0.30 \right ) \times
10^{-4} \, . 
\eeq
The experimental error is rapidly approaching the level of accuracy  
of the theoretical prediction. The main limiting factor on the
theoretical side resides in the perturbative QCD calculation and is
related to the ambiguity in the definition of the charm quark mass in
some two-loop diagrams containing the charm quark
\cite{gambino-misiak}. To improve significantly the present   
Next-to-Leading-Order (NLO) QCD calculation, one would need to include
one more order in the strong coupling expansion, and compute at least
the dominant NNLO effects: a  very challenging enterprise, which seems
to have already captured the imagination of some theorists
\cite{Bieri:2003ue}.  

The present calculation of the branching ratio of $\BtoXsgamma$
consists of several parts that are worth recalling. 
Perturbative QCD effects play an important role, due to the
presence of large logarithms $L = \ln m_b/\mw$, that can be resummed
using the formalism of the operator product expansion and
renormalization group techniques. The main components of the NLO 
calculation, which aims at resumming all the next-to-leading $\ord 
(\as^n L^{n-1})$ logarithms, have been established more than six years
ago. They are $i )$ the $\ord (\as)$ corrections to the relevant
Wilson coefficients \cite{wilson,Ciuchini:1998xe,BMU}, $ii )$ the
$\ord (\as)$ matrix elements of the corresponding 
dimension-five and six operators \cite{matrix}, and $iii )$ the $\ord
(\as^2)$ \AD \ Matrix (ADM) describing the mixing of physical
dimension-five and six operators
\cite{ADM4q,Chetyrkin:1998gb,misiak-munz,Chetyrkin:1997vx}. After the
$\ord (\as)$ matrix elements of some suppressed operators have  been
computed last year \cite{misiak-buras}, the NLO calculation is now
formally complete. Higher order electroweak \cite{electroweak,kagan}
and non-perturbative effects \cite{nonpert,nonpert2} amount to a few
percent in the total rate and seem to be well under control. However,
as the actual measurements adopt a lower cut on the photon energy --- 
typically around $2 \GeV$ --- the total rate is not accessible
experimentally. In fact, unless the photon energy cut lies well below 
$2 \GeV$, the measured rate  depends significantly on the
non-perturbative structure of the $B$ meson, and in particular on the
Fermi motion of the bottom quark inside the $B$ meson. The ensuing
theoretical uncertainty is significant \cite{kagan} and is generally
included in the reported experimental error, since the experimental
measurements are extrapolated to conventionally defined total rates.  

Nearly all the ingredients of the NLO QCD calculation involve a
considerable degree of technical sophistication and have been
performed independently by at least two groups, sometimes using
different methods. However, the most complex part of the whole
enterprise, the calculation of the two-loop dimension-five
\cite{misiak-munz} and of the three-loop dimension-six 
\cite{Chetyrkin:1997vx} $\ord (\as^2)$ ADM, has never been checked by 
a different group. The main purpose of this work is to present an
independent calculation of the $\ord (\as^2)$ ADM governing the
$\btosgamma$ and $\btosgluon$ transitions.  

The rare semileptonic decay $\BtoXslpluslminus$ represents, for new
physics searches, a route complementary to the radiative
ones. Experimentally, the exclusive modes have been recently  observed
for the first time \cite{expll}, and we now also have a measurement of
the inclusive branching fraction \cite{expllincl}. More progress is
expected in the future from the $B$ factories. Because of the presence
of large logarithms already at zeroth order in $\as$, a precise
calculation of the $\BtoXslpluslminus$ rate involves the resummation
of formally next-to-next-to-leading $\ord (\as^n L^{n - 2})$
logarithms. The Next-to-Next-to-Leading-Order (NNLO) QCD calculation
of $\BtoXslpluslminus$ has required  the computation of $i )$ the $\ord
(\as)$ corrections to  the corresponding Wilson coefficients \cite{BMU}
and $ii )$ the associated matrix elements at $\ord(\as)$
\cite{bsll,bsllgino}. Moreover, it involves $iii )$ the $\ord (\as^2)$
ADM, but the operator basis must be enlarged to include the
semileptonic operators characteristic of the $\btoslpluslminus$ 
transition. The $\ord (\as^2)$ mixing between four-quark and the
relevant semileptonic operators is one of the last missing pieces of a
full NNLO analysis of rare semileptonic $B$ meson decays. Since we
include the semileptonic operators in the $\ord (\as^2)$ calculation,
we are able to close this gap. Notice however, that a formally
complete NNLO calculation of the $\BtoXslpluslminus$ rate would also
require the knowledge of the $\ord (\as^2)$ self-mixing of the
four-quark operators \cite{inpreparation}, of the $\ord (\as)$ matrix
elements of the QCD penguin operators, and of the
renormalization-group-invariant two-loop matrix element of the 
vector-like semileptonic operator.   

We perform the calculation off-shell in an arbitrary $R_\xi$
gauge which allows us to explicitly check the gauge-parameter 
independence of the mixing among physical operators. To distinguish
between infrared (IR) and ultraviolet (UV) divergences we follow
\cite{Chetyrkin:1998fm} and introduce a common mass $M$ for all
fields, expanding all loop integrals in inverse powers of $M$. This
makes the calculation of the UV divergences possible even at three
loops, as $M$ becomes the only relevant internal scale and three-loop
tadpole integrals with a single non-zero mass are known. On the other
hand, this procedure requires  to take into account insertions of
non-physical operators, as well as of appropriate counterterms.   

We have so far emphasized the inclusive modes, as they are amenable to
a cleaner theoretical description. However, one should not
underestimate the importance  of the exclusive $B$ meson decays like 
$\BtoKstargamma$ \cite{bkstargammacleo}, $\BtoKstarlpluslminus$
\cite{expll}, $\Btorhogamma$
\cite{bkstargammacleo,brhogammaupperlimits} and   
$\Btorholpluslminus$. A thorough study of the exclusive channels can
yield useful additional information in testing the flavor sector of
the SM. These processes have received a lot of theoretical interest in
recent years and their accurate description will also benefit from a
firm understanding of higher order perturbative corrections.  

The ADM we have computed can be used in analyses of new physics
models, provided they do not introduce new operators with respect to
the SM. This applies, for example, to the case of two Higgs doublet
models
\cite{gambino-misiak,Ciuchini:1998xe,twohdm,bobeth-misiak-urban}, and
to some supersymmetric scenarios with minimal flavor violation. See 
for instance \cite{bobeth-misiak-urban,mfv}. On the other hand, in
left-right-symmetric models \cite{bobeth-misiak-urban,leftright} and
in the general minimal supersymmetric SM \cite{MSSM}, additional
operators with different chirality structures arise. In many cases one
can exploit the chiral invariance of QCD and use the same ADM, but 
in general an extended basis is required.

Our paper is organized as follows: in Section 2 we recall the relevant
effective Lagrangian and list all the dimension-five and six 
operators that will be needed in the calculation. In Section 3 we
review the renormalization procedure and explain how the operator 
renormalization matrix can be extracted from the matrix elements at
higher orders. The actual two- and three-loop calculation is described
in Section 4, while the results for the ADM are presented in Section 
5. We provide some intermediate results which can be useful in a 
number of related applications in an Appendix. They include the
complete $\ord (\as)$ and the relevant entries of the $\ord (\as^2)$
operator renormalization matrices.  

\section{The Effective Lagrangian} 

Let us briefly recall the formalism. We work in the framework of an 
effective low-energy theory with five active quarks, three active
leptons, photons and gluons, obtained by integrating out heavy degrees
of freedom characterized by a mass scale $M \ge \MW$. In the leading
order of the operator product expansion the effective off-shell
Lagrangian relevant for the $\btosgamma$, $\btosgluon$ and
$\btoslpluslminus$ transition at a scale $\mu$ is given by   
\beq \label{eq:effectivelagrangian}
\Leff = \LQCDQED + \f{4 \GF}{\sqrt{2}} V^\ast_{ts} V_{tb} \sum^{32}_{i
= 1} C_i (\mu) \, Q_i \, .     
\eeq
Here the first term is the conventional QCD-QED Lagrangian for the
light SM particles. In the second term $V_{ij}$ denotes the 
elements of the CKM matrix and $C_i (\mu)$ are the Wilson coefficients
of the corresponding operators $Q_i$ built out of the light fields. 

In our case it is useful to divide the local operators $Q_i$ entering
the effective Lagrangian into five different classes: $i )$ physical
operators, $ii )$ gauge-invariant operators that vanish by use of the
${\rm QCD} \hspace{-0.5mm} \times \hspace{-0.5mm} {\rm QED}$ Equations
Of Motion (EOM), $iii )$ gauge-variant EOM-vanishing operators, and
$iv )$ so-called evanescent operators that vanish algebraically in
four dimensions. In principle, one could  also encounter $v )$
non-physical counterterms that can be written as a
Becchi-Rouet-Stora-Tyutin (BRST) variation of some other operators,
so-called BRST-exact operators. However, they turn out to be
unnecessary in the case of the $\ord (\as^2)$ mixing of the operators
$Q_i$ considered below. See also \cite{misiak-munz}.      

Neglecting the mass of the strange quark and the electroweak penguin
operators which first arise at $\ord (\aem)$, the physical operators
can be written as \cite{Grinstein:1988pu}
\begin{align} \label{eq:physicaloperators} 
Q_1 & = (\bar{s}_L \gamma_\mu T^a c_L) (\bar{c}_L \gamma^\mu T^a b_L)
\, , \non \\
Q_2 & = (\bar{s}_L \gamma_\mu c_L) (\bar{c}_L \gamma^\mu b_L) \, ,
\non \\   
Q_3 & = (\bar{s}_L \gamma_\mu b_L) \sum\nolimits_q (\bar{q} \gamma^\mu
q) \, , \non \\ 
Q_4 & = (\bar{s}_L \gamma_\mu T^a b_L) \sum\nolimits_q (\bar{q}
\gamma^\mu T^a q) \, , \non \\ 
Q_5 & = (\bar{s}_L \gamma_\mu \gamma_\nu \gamma_\rho b_L)
\sum\nolimits_q (\bar{q} \gamma^\mu \gamma^\nu \gamma^\rho q) \, ,
\non \\   
Q_6 & = (\bar{s}_L \gamma_\mu \gamma_\nu \gamma_\rho T^a b_L)
\sum\nolimits_q (\bar{q} \gamma^\mu \gamma^\nu \gamma^\rho T^a q) \, ,
\non \\ 
Q_7 & = \f{e}{\gs^2} \mb (\bar{s}_L \sigma^{\mu \nu} b_R) F_{\mu \nu}
\, , \non \\  
Q_8 & = \f{1}{\gs} \mb (\bar{s}_L \sigma^{\mu \nu} T^a b_R) G_{\mu
  \nu}^a \, , \non \\   
Q_9 & = \f{e^2}{\gs^2} (\bar{s}_L \gamma_\mu b_L) \sum\nolimits_\ell  
(\bar{\ell} \gamma^\mu \ell) \, , \non \\   
Q_{10} & = \f{e^2}{\gs^2} (\bar{s}_L \gamma_\mu b_L)
\sum\nolimits_\ell (\bar{\ell} \gamma^\mu \gamma_5 \ell) \, ,      
\end{align}
where the sum over $q$ and $\ell$ extends over all light quark and
lepton fields, respectively. $e$ ($\gs$) is the electromagnetic
(strong) coupling constant, $q_L$ and $q_R$ are the chiral quark
fields, $F_{\mu \nu}$ ($G_{\mu \nu}^a$) is the electromagnetic
(gluonic) field strength tensor, and $T^a$ are the color matrices,
normalized so that $\mbox{Tr} (T^a T^b) = \delta^{ab}/2$. Notice that 
since QCD is flavor blind it is not necessary for our purposes  to
consider the analogues of $Q_1$ and $Q_2$ involving the up instead of
the charm quark.   

The physical operators given above consist of the current-current
operators $Q_1$ and $Q_2$, the QCD penguin operators $Q_3$--$Q_6$, the
magnetic moment type operators $Q_7$ and $Q_8$, and the semileptonic
operators $Q_9$ and $Q_{10}$, relevant for the $\btoslpluslminus$
transition. We have defined $Q_1$--$Q_{6}$ in such a way that problems
connected with the treatment of $\gamma_5$ in $\dimension = 4 - 2 \ep$
dimensions do not arise \cite{Chetyrkin:1998gb}. Consequently, we are
allowed to consistently use fully anticommuting $\gamma_5$ in
dimensional regularization throughout the calculation. 

The gauge-invariant EOM-vanishing operators can be chosen to be
\cite{Ciuchini:1998xe, BMU}  
\beq \label{eq:eomvanishinggaugeinvariantoperators}
\begin{split} 
Q_{11} & = \f{e}{\gs^2} \bar{s}_L \gamma^\mu b_L \partial^\nu F_{\mu 
  \nu} + \f{e^2}{\gs^2} (\bar{s}_L \gamma_\mu b_L) \sum\nolimits_f Q_f
  (\bar{f} \gamma^\mu f) \, , \\  
Q_{12} & = \f{1}{\gs} \bar{s}_L \gamma^\mu T^a b_L D^\nu G_{\mu \nu}^a
  + Q_4 \, , \\ 
Q_{13} & = \f{1}{\gs^2} \mb \bar{s}_L \Dsl \Dsl b_R \, , \\
Q_{14} & = \f{i}{\gs^2} \bar{s}_L \Dsl \Dsl \Dsl b_L \, , \\  
Q_{15} & = \f{i e}{\gs^2} \left [ \bar{s}_L
  \stackrel{\leftarrow}{\Dsl} \sigma^{\mu \nu} b_L F_{\mu \nu} - 
  F_{\mu \nu} \bar{s}_L \sigma^{\mu \nu} \Dsl b_L \right ] + Q_7 \, ,
  \\     
Q_{16} & = \f{i}{\gs} \left [ \bar{s}_L \stackrel{\leftarrow}{\Dsl}
  \sigma^{\mu \nu} T^a b_L G_{\mu \nu}^a - G_{\mu \nu}^a \bar{s}_L T^a
  \sigma^{\mu \nu} \Dsl b_L \right ] + Q_8 \, , 
\end{split}
\eeq
where the sum over $f$ runs over all light fermion fields, while 
$D_\mu$ and $\stackrel{\! \leftarrow}{D_\mu}$ denotes the covariant
derivative of the gauge group $\SUC \times \UQ$ acting on the fields
to the right and left, respectively. Notice that the set of operators 
$Q_1$--$Q_{16}$ closes off-shell under QCD renormalization, up to
evanescent operators \cite{Grinstein:1988pu,BMU}.  

In order to remove the divergences of all possible one-particle
irreducible (1PI) Green's functions with single insertion of
$Q_1$--$Q_{10}$ we also have to introduce the following gauge-variant
EOM-vanishing operators  
\beq \label{eq:eomvanishinggaugevariantoperators}
\begin{split}
Q_{17} & = \comment{-} \f{i}{\gs} \mb \bar{s}_L \left [
\stackrel{\leftarrow}{\Dsl} \Gsl - \Gsl \Dsl \right ] b_R \, , \\    
Q_{18} & = i \left [ \bar{s}_L \left ( \stackrel{\leftarrow}{\Dsl}
  \Gsl \Gsl - \Gsl \Gsl \Dsl \right ) b_L - i \mb \bar{s}_L \Gsl \Gsl
  b_R \right ] \, , \\ 
Q_{19} & = \comment{-} \f{1}{\gs} \left [ \bar{s}_L \left (
  \stackrel{\leftarrow}{\Dsl} \stackrel{\leftarrow}{\Dsl} \Gsl + \Gsl
  \Dsl \Dsl \right ) b_L + i \mb \bar{s}_L \Gsl \Dsl 
  b_R \right ] \, , \\
Q_{20} & = i \left [ \bar{s}_L \left ( \stackrel{\leftarrow}{\Dsl}
  G^a_\mu G^{a \mu} - G^a_\mu G^{a \mu} \Dsl \right ) b_L - i \mb
  \bar{s}_L G^a_\mu G^{a \mu} b_R \right ] \, , \\
Q_{21} & = \comment{-} \f{1}{\gs} \left [ \bar{s}_L \left (
  \stackrel{\leftarrow}{\Dsl} \stackrel{\leftarrow
  \hspace{1.5mm}}{D_\mu} G^\mu + G_\mu D^\mu \Dsl \right ) b_L + i \mb
  \bar{s}_L G_\mu D^\mu b_R \right ] \, , \\ 
Q_{22} & = \comment{-} \f{1}{\gs} \left [ \bar{s}_L \left (
  \stackrel{\leftarrow}{\Dsl} T^a + T^a \Dsl \right ) b_L +  i \mb
  \bar{s}_L T^a b_R \right ] \partial^\mu G^a_\mu \, , \\
Q_{23} & = \f{1}{\gs} \left [ \bar{s}_L \stackrel{\leftarrow}{\Dsl}
  \Gsl \Dsl b_L + i \mb \bar{s}_L \stackrel{\leftarrow}{\Dsl} \Gsl b_R 
  \right ] \, , \\
Q_{24} & = \comment{-} d^{abc} \left [ \bar{s}_L \left (
  \stackrel{\leftarrow}{\Dsl} T^a - T^a \Dsl \right ) b_L - i \mb
  \bar{s}_L T^a b_R \right ] G^b_\mu G^{c \mu} \, , 
\end{split}
\eeq
where $G^a_\mu$ denotes the  gluon field, and we have used the
abbreviations $G_\mu = G^a_\mu T^a$ and $d^{abc} = 2 \mbox{Tr} ( \{ 
T^a, T^b \} T^c )$.     

It is important to remark that the EOM-vanishing operators introduced
in
\Eqsand{eq:eomvanishinggaugeinvariantoperators}{eq:eomvanishinggaugevariantoperators}
arise as counterterms independently of what kind of IR regularization
is adopted in the computation. However, if the regularization respects
the underlying symmetry, and all the diagrams are calculated without
expansion in the external momenta, non-physical operators have
vanishing matrix elements \cite{theorems}. In this case the
EOM-vanishing operators given in
\Eqsand{eq:eomvanishinggaugeinvariantoperators}{eq:eomvanishinggaugevariantoperators}
play no role in the calculation of the mixing of physical operators. 
If the gauge symmetry is broken this is no longer the case, as
diagrams with insertions of non-physical operators will generally have
non-vanishing projection on the physical operators. Since our IR
regularization implies a massive gluon propagator, non-physical
counterterms play a crucial role at intermediate stages of the
calculation.      

As far as the evanescent operators are concerned, another eight 
operators are needed in order to find the $\ord (\alpha_s^2)$ mixing
of the physical operators $Q_1$--$Q_{10}$. Following
\cite{Chetyrkin:1997vx, Chetyrkin:1998gb} they can be defined as   
\begin{align} \label{eq:evanescentoperators}
Q_{25} & = (\bar{s}_L \gamma_\mu \gamma_\nu \gamma_\rho T^a c_L)
(\bar{c}_L \gamma^\mu \gamma^\nu \gamma^\rho T^a b_L) - 16 Q_1 \, 
, \non \\
Q_{26} & = (\bar{s}_L \gamma_\mu \gamma_\nu \gamma_\rho c_L) 
(\bar{c}_L \gamma^\mu \gamma^\nu \gamma^\rho b_L) - 16 Q_2 \, , \non \\ 
Q_{27} & = (\bar{s}_L \gamma_\mu \gamma_\nu \gamma_\rho \gamma_\sigma
\gamma_\tau b_L) \sum\nolimits_q (\bar{q} \gamma^\mu \gamma^\nu 
\gamma^\rho \gamma^\sigma \gamma^\tau q) + 64 Q_3 - 20 Q_5 \, , \non \\
Q_{28} & = (\bar{s}_L \gamma_\mu \gamma_\nu \gamma_\rho \gamma_\sigma 
\gamma_\tau T^a b_L) \sum\nolimits_q (\bar{q} \gamma^\mu \gamma^\nu
\gamma^\rho \gamma^\sigma \gamma^\tau T^a q) + 64 Q_4 - 20 Q_6 \, ,
\non \\
Q_{29} & = (\bar{s}_L \gamma_\mu \gamma_\nu \gamma_\rho \gamma_\sigma
\gamma_\tau T^a c_L) (\bar{c}_L \gamma^\mu \gamma^\nu \gamma^\rho
\gamma^\sigma \gamma^\tau T^a b_L) - 256 Q_1 - 20 Q_{25} \, , \non \\ 
Q_{30} & = (\bar{s}_L \gamma_\mu \gamma_\nu \gamma_\rho \gamma_\sigma
\gamma_\tau c_L) (\bar{c}_L \gamma^\mu \gamma^\nu \gamma^\rho
\gamma^\sigma \gamma^\tau b_L) - 256 Q_2 - 20 Q_{26} \, , \non \\  
Q_{31} & = (\bar{s}_L \gamma_\mu \gamma_\nu \gamma_\rho \gamma_\sigma
\gamma_\tau \gamma_\upsilon \gamma_\omega b_L) \sum\nolimits_q
(\bar{q} \gamma^\mu \gamma^\nu \gamma^\rho \gamma^\sigma \gamma^\tau 
\gamma^\upsilon \gamma^\omega q) + 1280 Q_3 - 336 Q_5 \, , \non \\
Q_{32} & = (\bar{s}_L \gamma_\mu \gamma_\nu \gamma_\rho \gamma_\sigma
\gamma_\tau \gamma_\upsilon \gamma_\omega T^a b_L) \sum\nolimits_q 
(\bar{q} \gamma^\mu \gamma^\nu \gamma^\rho \gamma^\sigma \gamma^\tau
\gamma^\upsilon \gamma^\omega T^a q) + 1280 Q_4 - 336 Q_6 \, . 
\end{align}
  
\section{Renormalization of the Effective Theory}

Our aim is to study the renormalization properties of the physical
operators $Q_1$--$Q_{10}$ introduced in
\Eq{eq:physicaloperators}. Upon renormalization, the bare Wilson 
coefficients $C_{i, B} (\mu)$ of \Eq{eq:effectivelagrangian} transform
as    
\beq \label{eq:effectivebarelag}
C_{i, B} (\mu) = Z_{ji} C_j (\mu) \, , 
\eeq 
where the renormalization constants $Z_{ij}$ can be expanded in powers
of $\as$ as  
\beq \label{eq:zfactors}
Z_{ij} = \delta_{ij} + \sum^\infty_{k = 1} \left ( \f{\as}{4 \pi}
\right )^k Z_{ij}^{(k)} \, , \hspace{5mm} \text{with} \hspace{5mm}
Z_{ij}^{(k)} = \sum_{l = 0}^k \frac{1}{\eps^l} Z_{ij}^{(k,l)} \, .   
\eeq
Following the standard $\MSbar$ scheme prescription, $Z_{ij}$ is given
by pure $1/\eps^l$ poles, except when $i = 25$--$32$ and $j \neq
25$--$32$. In the latter case, the renormalization constant is finite,
to make sure that the matrix elements of the evanescent operators
vanish in four dimensions \cite{Buras:1990xd,evanescent}. The
calculation of an effective amplitude $\Aeff$, also involves the
matrix element $\langle Q_i \rangle \equiv \langle F | Q_i (\mu) | I
\rangle$ of the operator $Q_i$ between a initial state $I$ and a final
state $F$, which is renormalized by the usual coupling, mass and wave
function renormalization factor characteristic of the operator, $Q_i
\to Z(Q_i)$. The renormalized effective amplitude is therefore given
by 
\beq \label{eq:fullmatrixelementrenormalization} 
\Aeff = Z_{ji} C_j (\mu) \langle Z(Q_i) \rangle_{R} \, ,
\eeq
where $\langle Z(Q_i) \rangle_{R}$ denotes the matrix element of the
operator $Z(Q_i)$ after performing coupling, mass and wave function
renormalization. Clearly, it is also possible to define the operator
renormalization constant $\overline{\! Z}_{ij}$ from the relation
 between unrenormalized and amputated Green's functions via $\langle
Z(Q_i) \rangle_{R} = \overline{\! Z}_{ij} \langle Q_j \rangle_{B}$. In
this case, one simply has $\overline{\! Z}_{ij} = {Z}_{ij}^{-1}$.    
In general $Z(Q_i)$ will not be proportional to $Q_i$. For example, in
many of the EOM-vanishing operator introduced in 
\Eqsand{eq:eomvanishinggaugeinvariantoperators}{eq:eomvanishinggaugevariantoperators}
one has two different terms, only one of which has a factor of
$\mb$. Correspondingly, the $\mb$ renormalization of the operator is  
\beq \label{eq:mboperators}
Z_{\mb} (Q_i) = Q_i + ( Z_{\mb} - 1 ) Q^\prime_i \, , 
\eeq
where $Z_{\mb}$ denotes the mass renormalization constant of the
bottom quark, and $Q^\prime_i$ is the part of $Q_i$ proportional to
$\mb$.   

{%
\begin{figure}[t]
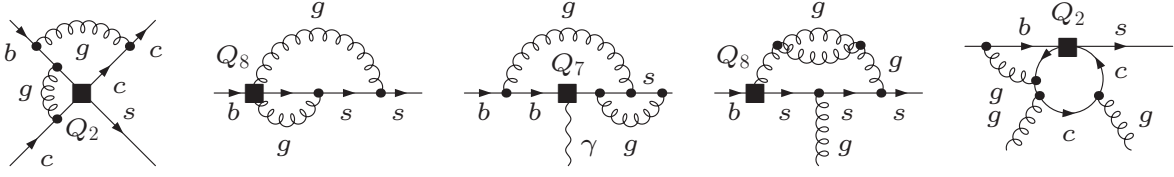

\begin{center}
\scalebox{1.6}{\begin{picture}(55,45)(0,-5)\input{bscc.tex}\end{picture}}%
\hspace{-1mm}
\scalebox{1.6}{\begin{picture}(55,45)(0,-5)\input{bs.tex}\end{picture}}%
\hspace{1mm}
\scalebox{1.6}{\begin{picture}(55,45)(0,-5)\input{bsa.tex}\end{picture}}%
\hspace{1mm}
\scalebox{1.6}{\begin{picture}(55,45)(0,-5)\input{bsg.tex}\end{picture}}%
\hspace{1mm}
\scalebox{1.6}{\begin{picture}(55,45)(0,-5)\input{bsgg.tex}\end{picture}}%
\caption{ Some of the two-loop 1PI diagrams we had to
calculate in order to find the $\ord (\as^2)$ mixing  of the
complete set of operators $Q_1$--$Q_{32}$.}
\label{fig:twoloop} 
\end{center}
\end{figure}
}%

The product on the right-hand side of
\Eq{eq:fullmatrixelementrenormalization} must be finite by definition
at any given order in $\as$. Therefore, requiring the cancellation of
UV divergences we can extract $Z_{ij}^{(k)}$ order by order. The
result, up to third order, reads   
\beq \label{eq:extractionzijrenormalized}
\begin{split}
Z_{ij}^{(1)} \langle Q_j \rangle^{(0)}_{R} & = -\langle
Z(Q_i) \rangle^{(1)}_{R} \, , \\
Z_{ij}^{(2)} \langle Q_j \rangle^{(0)}_{R} & = -\langle Z(Q_i)
\rangle^{(2)}_{R} - Z_{ij}^{(1)} \langle Z(Q_j) \rangle^{(1)}_{R} \, ,
\\
Z_{ij}^{(3)} \langle Q_j \rangle^{(0)}_{R} & = -\langle Z(Q_i)
\rangle^{(3)}_{R} - Z_{ij}^{(1)} \langle Z(Q_j) \rangle^{(2)}_{R} - 
Z_{ij}^{(2)} \langle Z(Q_j) \rangle^{(1)}_{R} \, , 
\end{split}
\eeq
where the superscript $(k)$ always stands for the $k$-th order
contribution in $\as$. 

If we leave aside the complication that in general $Z(Q_i)$ will not
be proportional to $Q_i$, and write symbolically $\langle Z (Q_i)
\rangle_{R} = Z_i \langle Q_i \rangle_{B}$ the above relations can be
rewritten in terms of bare quantities. Up to third order in $\as$ we
obtain   
\beq \label{eq:extractionzijbare}
\begin{split}
Z_{ij}^{(1)} \langle Q_j \rangle^{(0)}_{B} & = -\langle
Q_i \rangle^{(1)}_{B} - Z_i \langle Q_i
\rangle^{(0)}_{B} \, , \\ 
Z_{ij}^{(2)} \langle Q_j \rangle^{(0)}_{B} & = -\langle
Q_i \rangle^{(2)}_{B} - Z^{(1)}_{ij} \langle Q_j
\rangle^{(1)}_{B} - Z^{(1)}_i \langle Q_i
\rangle^{(1)}_{B} \\  
& - Z^{(1)}_{ij} Z^{(1)}_j \langle Q_j \rangle^{(0)}_{B} -
Z^{(2)}_i \langle Q_i \rangle^{(0)}_{B} \, , \\     
Z_{ij}^{(3)} \langle Q_j \rangle^{(0)}_{B} & = -\langle
Q_i \rangle^{(3)}_{B} - Z^{(1)}_{ij} \langle Q_j
\rangle^{(2)}_{B} - Z^{(1)}_i \langle Q_i
\rangle^{(2)}_{B} \\ 
& -Z^{(2)}_{ij} \langle Q_j \rangle^{(1)}_{B} -
Z^{(1)}_{ij} Z^{(1)}_j \langle Q_j \rangle^{(1)}_{B} -
Z^{(2)}_i \langle Q_i \rangle^{(1)}_{B} \\  
& -Z^{(2)}_{ij} Z^{(1)}_j \langle Q_j \rangle^{(0)}_{B} -
Z^{(1)}_{ij} Z^{(2)}_j \langle Q_j \rangle^{(0)}_{B} -
Z^{(3)}_i \langle Q_i \rangle^{(0)}_{B} \, .  
\end{split}
\eeq
The first line in \Eqs{eq:extractionzijbare} recalls the familiar
result that the one-loop renormalization matrix is given by the UV
divergences of the one-loop matrix elements, after performing wave
function and possibly coupling and mass renormalization. For example,
in the case of the operators $Q_1$--$Q_6$, one has $Z_i = Z_q^2$ 
with $Z_q$ denoting the wave function renormalization constant
of the quark fields, and \Eqs{eq:extractionzijbare} take a
particularly simple form, which upon expansion in $\as$ reproduces the
classical results derived more than ten years ago \cite{Buras:1990xd}.

{%
\begin{figure}[t]
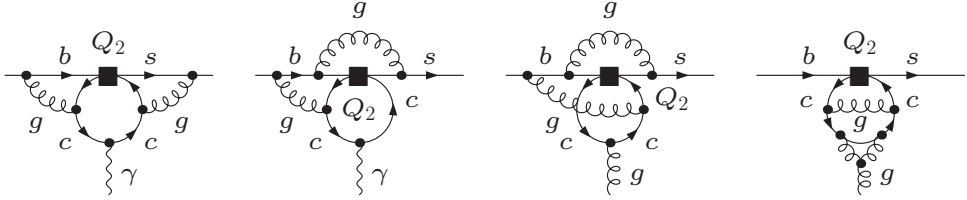

\begin{center}
\scalebox{1.6}{\begin{picture}(55,45)(0,-5)\input{penguin1.tex}\end{picture}}%
\hspace{1mm}
\scalebox{1.6}{\begin{picture}(55,45)(0,-5)\input{penguin2.tex}\end{picture}}%
\hspace{1mm}
\scalebox{1.6}{\begin{picture}(55,45)(0,-5)\input{penguin3.tex}\end{picture}}%
\hspace{1mm}
\scalebox{1.6}{\begin{picture}(55,45)(0,-5)\input{penguin4.tex}\end{picture}}%
\caption{ Some of the three-loop 1PI diagrams we had to
calculate in order to find the mixing of the four-quark operators
into $Q_7$--$Q_{10}$ at $\ord (\as^3)$.}  
\label{fig:threeloop}
\end{center}
\end{figure}
}%

For a given set of operators and knowing the QCD renormalization
constants, the solution of the above systems of linear equations
requires the calculation of a sufficient number of Green's functions
for different external fields with single insertions of the operators
$Q_i$. In our case, in order to determine the complete $Z_{ij}^{(k)}$
of all the operators introduced in Section 2, it is sufficient to
calculate the $\ord (\as^k)$ matrix elements of $Q_1$--$Q_{32}$ for
the $\btosccbar$, $\btos$, $\btosgamma$, $\btosgluon$ and
$\btosgluongluon$ transitions --- see \Fig{fig:twoloop}. As
we are interested in a subset of the three-loop ADM, we have actually
calculated only the three-loop $\btosgamma$ and $\btosgluon$
amplitudes involving insertions of $Q_1$--$Q_6$ --- see
\Fig{fig:threeloop}. We have calculated the complete off-shell
amplitudes up to terms proportional to external momenta squared. By
using the EOM it is therefore straightforward to extract the mixing
into $Q_7$--$Q_{10}$. Notice that the results for the $Z_{ij}^{(k)}$
cannot depend on the considered Green's functions and that the pole
parts need to have the structure of the complete set of local
operators $Q_1$--$Q_{32}$. Both features represent a powerful
consistency check of the computation of the renormalization constants 
$Z_{ij}^{(k)}$.           

The normalization of the physical operators adopted in Section 2 has
been chosen \cite{BMU} in such a way that the power of $\as$ in
$Z_{ij}$ is equal to the number of loops of the contributing
diagrams. For instance, without the factor $1/\gs^2$ in
$Q_7$--$Q_{10}$, as in the standard normalization 
adopted in \cite{gambino-misiak,misiak-buras,bsllgino}, both one- and
two-loop diagrams contribute to the $\ord (\as)$ mixing matrix,
because of the $\ord (\as)$ two-loop mixing of four-quark into
magnetic operators. This choice simplifies both the implementation of
the renormalization program and the resummation of large logarithms,
since the redefinition enables one to proceed for $\btoslpluslminus$
in the same way as in the $\btosgamma$ and $\btosgluon$ case. 

In a mass independent renormalization scheme $Z_{ij}$ is
$\mu$-independent. This allows to check the renormalization of two-
and three-loop matrix elements. The right-hand sides of the  
\Eqsand{eq:extractionzijrenormalized}{eq:extractionzijbare} receive
contributions from irreducible two- and three-loop diagrams as well as
one- and two-loop counterterms. The $\mu$-dependence is different in
each case and governed by the $n$-loop factor $\left ( \mu^{2 \eps}
\right )^n$. The UV structure of the $k$-th term is therefore given by
\beq \label{eq:uvstructure}
Z_{ij}^{(k)} \langle Q_j \rangle_{B}^{(0)} = \sum_{n = 0}^k \sum_{l =
1}^{n} \left ( \mu^{2 \eps} \right )^n \f{1}{\eps^l} M^{(n,l)} \, , 
\end{equation}
where $M^{(n,l)}$ denotes the $1/\ep^l$ pole of the sum of all
$n$-loop contributions. Expanding in  powers of $\eps$ we find 
the following set of equations which have to be fulfilled to get a
$\mu$-independent $Z_{ij}$ up to three-loop order:  
\begin{align} \label{eq:muindependence}
3 M^{(3,2)} + 2 M^{(2,2)} + M^{(1,2)} & = 0 \, , \non \\
3 M^{(3,3)} + 2 M^{(2,3)} + M^{(1,3)} & = 0 \, , \non \\
9 M^{(3,3)} + 4 M^{(2,3)} + M^{(1,3)} & = 0 \, .  
\end{align}
This system of equations provides us with a powerful check of the
renormalization of two- as well as three-loop diagrams. Notice that
the locality of UV divergences also places some constraints on the 
renormalization matrix itself. We will turn to this point later on. 

\section{The Calculation of the Operator Mixing}

In the renormalization of QCD and QED at higher orders the standard
method of extracting the UV divergence structure of a Feynman integral
is to perform the calculation with massless propagators. However, if
one uses massless propagators to compute three-point or higher Green's
functions one might generate spurious IR infinities which, in
dimensional regularization, cannot be distinguished from the UV
divergences one seeks. There exist several methods
\cite{irrearrangment} to overcome this problem, but they are generally
quite involved and not suitable to the automated evaluation of a large
number of diagrams. In the  approach of \cite{Chetyrkin:1998fm} the
so-called IR rearrangement is performed by introducing an artificial
mass. For the calculation of the renormalization constants this means
that we can safely apply Taylor expansion in the external momenta
after introducing a non-zero auxiliary mass $M$ for each internal
propagator, including those of the massless vector particles. The
auxiliary mass regulates all IR divergences and the renormalization
constants can be extracted from the UV divergences of massive,
one-scale tadpole diagrams, that are known up to the four-loop level
\cite{tadpoles}.   

Following references \cite{misiak-munz, Chetyrkin:1998fm}, the
starting point of our procedure is the exact decomposition of a 
propagator:   
\beq \label{eq:propagatordecomposition}
\f{1}{\left ( k + p \right )^2 - m^2} = \f{1}{k^2 - M^2} - \f{p^2 + 2
k \cdot p - m^2 + M^2}{k^2 - M^2} \f{1}{\left ( k + p \right )^2 -
m^2} \, .  
\eeq 
Here $k$ is a linear combination of the integration momenta, $p$
stands for a linear combination of the external momenta, and $m$ 
denotes the mass of the propagating particle. If we assume that the
dimensionality of the operators in our effective theory is bounded
from above, and we apply recursively the above decomposition a
sufficient number of times, we will reach the point where the overall 
degree of divergence of a certain  diagram would become negative if
any of its propagators were replaced by the last term in the
decomposition. We are then allowed to drop the last term in the 
propagator decomposition, as it does not affect the UV divergent part
of the Green's function after subtraction of all subdivergences. 

This algorithm can be also simplified by the following observation
\cite{Chetyrkin:1998fm}. The terms containing powers of the auxiliary
mass squared in the numerators contribute only to UV divergences that
are proportional to those powers of $M^2$. The latter are local after
the subtraction of all subdivergences, and must precisely cancel
similar terms originating from integrals with no auxiliary mass in the
numerators. Since the decomposition of \Eq{eq:propagatordecomposition} 
is exact, no dependence on $M^2$ can remain after performing the whole
calculation. This observation allows one to avoid calculating
integrals that contain an artificial mass in the numerator. Instead of
calculating them, one can just replace them by local counterterms
proportional to $M^2$ which cancel the corresponding subdivergences in
the integrals with no $M^2$ in the propagator
numerators. Nevertheless, the final result for the UV  divergent parts
of the Green's functions are precisely the same as if the full
propagators were used.    
  
The counterterms proportional to $M^2$ in general do not preserve the
symmetry of the underlying theory, specifically they do not have to
be gauge-invariant. Fortunately, the number of these counterterms is
usually rather small, because their dimension must be two units less
than the maximal dimension of the operators belonging to the effective
theory. For instance, in QCD only a single possible gauge-variant
operator exists that fulfills the above requirement. It looks like a
gluon mass counterterm,  
\beq \label{eq:gluonmass}
\mir^2 G_\mu^a G^{a \mu} \, , 
\eeq
and cancels gauge-variant pieces of integrals with no $M^2$ in the
numerators. To ensure that our renormalization procedure with the
fictitious gluon and photon mass is valid, we have checked explicitly
the full $\MSbar$ renormalization of QCD and QED up to the three-loop
level, finding perfect agreement with the results given in the
literature \cite{qcdqedrenormalization}. In our case, beside the term
in \Eq{eq:gluonmass}, we also have $M^2$ counterterms of
dimension-three and four, some of which explicitly break gauge
invariance:     
\beq \label{eq:gaugevariantoperators}
\f{\mir^2}{\gs^2} \mb \bar{s}_L b_R \, , \hspace{5mm} \f{i
\mir^2}{\gs^2} \bar{s}_L \dsl b_L \, , \hspace{5mm} \f{\mir^2
e}{\gs^2} \bar{s}_L \Asl b_L \, , \hspace{5mm} \f{\mir^2}{\gs}
\bar{s}_L \Gsl b_L \, ,   
\eeq
where $A_\mu$ denotes the  photon field. 
  
As already mentioned in Section 2, another side effect of our IR
regularization is that we have to consider insertions of non-physical
effective operators in our calculation. Let us explain this point in
more detail. Non-physical counterterms generally arise in QCD 
calculations, but the projections of their matrix elements on physical
operators vanish unless the underlying symmetry is broken at some
stage. Due to the exact nature of the decomposition
\Eq{eq:propagatordecomposition}, the UV poles of the diagrams obtained
by our method are correct after the subtraction of all
subdivergences. However, the UV poles related to subdivergences and
their subtraction terms both depend on the finite parts of certain 
lower loop diagrams, which in our approach are not necessarily correct
and do not comply with the usual Slavnov-Taylor identities. For
instance, the introduction of the IR regulator invalidates the 
argument that guarantees vanishing on-shell matrix elements for the
non-physical operators. One therefore expects non-negligible
contributions to the counterterms from all possible operators with
appropriate dimension. Consequently, all EOM-vanishing operators,  
gauge-invariant or not, and in general even BRST-exact operators 
must be included in the operator basis. The ``incorrect''
subdivergences are present in both counterterm and irreducible
diagrams, but they cancel in their sum, provided the calculation is
carried out in exactly the same way. The operator renormalization
constants calculated in this way are correct for all the operators in
the complete basis.  

The large number of diagrams which occurs at higher orders makes it
necessary to generate the diagrams automatically. For the evaluation
of the ADM  presented here all diagrams have been generated by the
{\sc Mathematica} \cite{mathematica} package {\it FeynArts}
\cite{feynarts}, which provides the possibility to implement the 
Feynman rules for different Lagrangians in a simple way. We have
adapted it to include the effective vertices induced by the operators
$Q_1$--$Q_{32}$. We have processed the {\it FeynArts} output using two
independent programs. In one case the output is converted into a
format recognizable by the language {\sc Form} \cite{Form}. The group
theory for each graph as well as the projection onto all possible form
factors is performed before the integrals are evaluated. The very
computation of  the integrals is done with the program package {\tt
MATAD} \cite{Steinhauser:2000ry}, which is able to deal with vacuum
diagrams at one-, two- and three-loop level where several of the
internal lines may have a common mass. The calculation of the tadpole
integrals in {\tt MATAD} is based on the so-called
integration-by-parts technique \cite{Tkachov:1981wb}. The second
program is entirely a {\sc Mathematica} code, which for the three-loop
integrals uses the algorithm described in detail in
\cite{Chetyrkin:1998fm}.     

\section{The \AD \ Matrix}

The \ads \ $\gamma_{ij}$ defined by 
\beq \label{eq:rge}
\mu \f{d}{d \mu} C_i (\mu) = \gamma_{ji} C_j (\mu) \, ,
\eeq 
can be expressed in terms of the entries of the renormalization matrix
$Z_{ij}$ as follows   
\beq \label{eq:defineanomalousdimmatrix}
\gamma_{ij} = Z_{ik} \hspace{0.5mm} \mu \f{d}{d \mu} Z^{-1}_{kj} \, . 
\eeq
In a mass independent renormalization scheme the only $\mu$-dependence
of $Z_{ij}$ resides in the coupling constant. In consequence, we might
rewrite \Eq{eq:defineanomalousdimmatrix} as 
\beq \label{eq:gammaintermsofbeta}
\gamma_{ij} = 2 \beta (\eps, \as) Z_{ik} \f{d}{d \as} Z^{-1}_{kj} \, ,
\eeq
where $\beta (\eps, \as) $ is related to the $\beta$ function
via     
\beq \label{eq:betaeps}
\beta (\eps, \as) = \as \big ( -\eps + \beta (\as) \big ) \, .    
\eeq
The finite parts of \Eq{eq:gammaintermsofbeta} in the limit of $\eps$
going to zero give the \ads. Expanding the \ads \ and the $\beta$
function in powers of $\as$ as     
\beq \label{eq:expansionbetagamma}
\hat{\gamma} = \sum^\infty_{k = 1} \left ( \f{\as}{4 \pi} \right )^k
\hat{\gamma}^{(k - 1)} \, , \hspace{5mm} \text{and} \hspace{5mm} \beta
(\as) = -\sum^\infty_{k = 1} \left ( \f{\as}{4 \pi} \right )^k
\beta_{k - 1} \, ,   
\eeq 
we find in accordance with \cite{Chetyrkin:1998fm} up to third order
in $\as$: 
\beq \label{eq:gammaexpansion}
\begin{split}
\hat{\gamma}^{(0)} & = 2 \hat{Z}^{(1,1)} \, , \\
\hat{\gamma}^{(1)} & = 4 \hat{Z}^{(2,1)} - 2 \hat{Z}^{(1,1)}
\hat{Z}^{(1,0)} - 2 \hat{Z}^{(1,0)} \hat{Z}^{(1,1)} + 2 \beta_0
\hat{Z}^{(1,0)} \, , \\ 
\hat{\gamma}^{(2)} & = 6 \hat{Z}^{(3,1)} - 4 \hat{Z}^{(2,1)}
\hat{Z}^{(1,0)} - 2 \hat{Z}^{(1,1)} \hat{Z}^{(2,0)} - 4
\hat{Z}^{(2,0)} \hat{Z}^{(1,1)} - 2 \hat{Z}^{(1,0)} \hat{Z}^{(2,1)} 
\\  
& + 2 \hat{Z}^{(1,1)} \hat{Z}^{(1,0)} \hat{Z}^{(1,0)} + 2
\hat{Z}^{(1,0)} \hat{Z}^{(1,1)} \hat{Z}^{(1,0)} + 2 \hat{Z}^{(1,0)} 
\hat{Z}^{(1,0)} \hat{Z}^{(1,1)} \\ 
& + 2 \beta_1 \hat{Z}^{(1,0)} + 4 \beta_0 \hat{Z}^{(2,0)} { -} 2
\beta_0 \hat{Z}^{(1,0)} \hat{Z}^{(1,0)} \, .      
\end{split}
\eeq
On the other hand the pole parts of \Eq{eq:gammaintermsofbeta} must
vanish. From this condition one obtains relations between single,
double and triple $1/\eps$ poles of the $Z_{ij}$, which constitute a 
useful check of the calculation. In agreement with
\cite{Chetyrkin:1998fm} we find  
\beq \label{eq:localitychecks}
\begin{split}
\hat{Z}^{(2,2)} & = \f{1}{2} \hat{Z}^{(1,1)} \hat{Z}^{(1,1)} -
\f{1}{2} \beta_0 \hat{Z}^{(1,1)} \, , \\ 
\hat{Z}^{(3,3)} & = \f{1}{6} \hat{Z}^{(1,1)} \hat{Z}^{(1,1)}
\hat{Z}^{(1,1)} - \f{1}{2} \beta_0 \hat{Z}^{(1,1)} \hat{Z}^{(1,1)} +
\f{1}{3} \beta_0^2 \hat{Z}^{(1,1)} \, , \\   
\hat{Z}^{(3,2)} & = \f{2}{3} \hat{Z}^{(2,1)} \hat{Z}^{(1,1)} 
+ \f{1}{3} \hat{Z}^{(1,1)} \hat{Z}^{(2,1)} - \f{1}{3} \hat{Z}^{(1,1)}
\hat{Z}^{(1,0)} \hat{Z}^{(1,1)} - \f{1}{6} \hat{Z}^{(1,0)}
\hat{Z}^{(1,1)} \hat{Z}^{(1,1)} \\ 
& - \f{1}{3} \beta_1 \hat{Z}^{(1,1)} - \f{2}{3} \beta_0
\hat{Z}^{(2,1)} + \f{1}{6} \beta_0 \hat{Z}^{(1,0)} \hat{Z}^{(1,1)} \,
.      
\end{split}
\eeq

Having summarized the general formalism and our method, we will now
present our results for five active quark flavors. For completeness we
start with the regularization- and renormalization-scheme independent
matrix $\hat{\gamma}^{(0)}$, which is given by    
\beq \label{eq:gamma0new}
\hat{\gamma}^{(0)} = 
\left (
\begin{array}{cccccccccccc}
-4 & {\scriptstyle \f{8}{3}} & 0 & {\scriptstyle -\f{2}{9}} & 0 & 0 &
0 & 0 & {\scriptstyle -\f{32}{27}} & 0 \\  
12 & 0 & 0 & {\scriptstyle \f{4}{3}} & 0 & 0 & 0 & 0 & {\scriptstyle
-\f{8}{9}} & 0 \\  
0 & 0 & 0 & {\scriptstyle -\f{52}{3}} & 0 & 2 & 0 & 0 & {\scriptstyle
-\f{16}{9}} & 0 \\  
0 & 0 & {\scriptstyle -\f{40}{9}} & {\scriptstyle -\f{100}{9}} &
{\scriptstyle \f{4}{9}} & {\scriptstyle \f{5}{6}} & 0 & 0 &
{\scriptstyle \f{32}{27}} & 0 \\  
0 & 0 & 0 & {\scriptstyle -\f{256}{3}} & 0 & 20 & 0 & 0 &
{\scriptstyle -\f{112}{9}} & 0 \\  
0 & 0 & {\scriptstyle -\f{256}{9}} & {\scriptstyle \f{56}{9}} &
{\scriptstyle \f{40}{9}} & {\scriptstyle -\f{2}{3}} & 0 & 0 &
{\scriptstyle \f{512}{27}} & 0 \\  
0 & 0 & 0 & 0 & 0 & 0 & {\scriptstyle -\f{14}{3}} & 0 & 0 & 0 \\ 
0 & 0 & 0 & 0 & 0 & 0 & {\scriptstyle -\f{32}{9}} & -6 & 0 & 0 \\ 
0 & 0 & 0 & 0 & 0 & 0 & 0 & 0 & {\scriptstyle -\f{46}{3}} & 0 \\ 
0 & 0 & 0 & 0 & 0 & 0 & 0 & 0 & 0 & {\scriptstyle -\f{46}{3}}
\end{array}
\right ) \, .  
\eeq
While the matrix $\hat{\gamma}^{(0)}$ is renormalization-scheme
independent, $\hat{\gamma}^{(1)}$ and $\hat{\gamma}^{(2)}$ are not. In
the $\MSbar$ scheme supplemented by the definition of evanescent
operators given in \Eq{eq:evanescentoperators} we obtain       
{%
\renewcommand{\arraycolsep}{3.5pt}
\beq \label{eq:gamma1new}
\hat{\gamma}^{(1)} = 
\left (
\begin{array}{cccccccccccc}
{\scriptstyle -\f{355}{9}} & {\scriptstyle -\f{502}{27}} &
{\scriptstyle -\f{1412}{243}} & {\scriptstyle -\f{1369}{243}} &
{\scriptstyle \f{134}{243}} & {\scriptstyle -\f{35}{162}} &
{\scriptstyle -\f{232}{243}} & {\scriptstyle \f{167}{162}} &
{\scriptstyle -\f{2272}{729}} & 0 \\  
 {\scriptstyle -\f{35}{3}} & {\scriptstyle -\f{28}{3}} & {\scriptstyle
-\f{416}{81}} & {\scriptstyle \f{1280}{81}} & {\scriptstyle
\f{56}{81}} & {\scriptstyle \f{35}{27}} & {\scriptstyle \f{464}{81}} &
{\scriptstyle \f{76}{27}} & {\scriptstyle \f{1952}{243}} & 0 \\  
0 & 0 & {\scriptstyle -\f{4468}{81}} & {\scriptstyle -\f{31469}{81}}
& {\scriptstyle \f{400}{81}} & {\scriptstyle \f{3373}{108}} &
{\scriptstyle \f{64}{81}} & {\scriptstyle \f{368}{27}} & {\scriptstyle
-\f{6752}{243}} & 0 \\  
0 & 0 & {\scriptstyle -\f{8158}{243}} & {\scriptstyle
-\f{59399}{243}} & {\scriptstyle \f{269}{486}} & {\scriptstyle
\f{12899}{648}} & {\scriptstyle -\f{200}{243}} & {\scriptstyle
-\f{1409}{162}} & {\scriptstyle -\f{2192}{729}} & 0 \\  
0 & 0 & {\scriptstyle -\f{251680}{81}} & {\scriptstyle
-\f{128648}{81}} & {\scriptstyle \f{23836}{81}} & {\scriptstyle
\f{6106}{27}} & {\scriptstyle -\f{6464}{81}} & {\scriptstyle
\f{13052}{27}} & {\scriptstyle -\f{84032}{243}} & 0 \\  
0 & 0 & {\scriptstyle \f{58640}{243}} & {\scriptstyle
-\f{26348}{243}} & {\scriptstyle -\f{14324}{243}} & {\scriptstyle
-\f{2551}{162}} & {\scriptstyle -\f{11408}{243}} & {\scriptstyle
-\f{2740}{81}} & {\scriptstyle -\f{37856}{729}} & 0 \\  
0 & 0 & 0 & 0 & 0 & 0 & {\scriptstyle \f{2600}{27}} & 0 & 0 & 0 \\  
0 & 0 & 0 & 0 & 0 & 0 & {\scriptstyle -\f{2192}{81}} & {\scriptstyle
\f{1975}{27}} & 0 & 0 \\  
0 & 0 & 0 & 0 & 0 & 0 & 0 & 0 & {\scriptstyle -\f{232}{3}} & 0 \\  
0 & 0 & 0 & 0 & 0 & 0 & 0 & 0 & 0 & {\scriptstyle -\f{232}{3}}
\end{array}
\right ) \, ,  
\eeq
}%
and 
\beq \label{eq:gamma2new}
\hat{\gamma}^{(2)} = 
\left (
\begin{array}{cccccccccccc}
\text{?} & \text{?} & \text{?} & \text{?} & \text{?} & \text{?} &
{\scriptstyle -\f{13234}{2187}} & {\scriptstyle \f{13957}{2916}} &
{\scriptstyle -\f{1359190}{19683}} {\scriptstyle +\f{6976}{243}
\zetathree} & 0 \\    
\text{?} & \text{?} & \text{?} & \text{?} & \text{?} & \text{?} &
{\scriptstyle \f{20204}{729}} & {\scriptstyle \f{14881}{972}} &
{\scriptstyle -\f{229696}{6561}} {\scriptstyle -\f{3584}{81}
\zetathree} & 0 \\   
0 & 0 & \text{?} & \text{?} & \text{?} & \text{?} & {\scriptstyle
\f{92224}{729}} & {\scriptstyle \f{66068}{243}} & {\scriptstyle
-\f{1290092}{6561}} {\scriptstyle +\f{3200}{81} \zetathree} & 0 \\   
0 & 0 & \text{?} & \text{?} & \text{?} & \text{?} & {\scriptstyle
-\f{184190}{2187}} & {\scriptstyle -\f{1417901}{5832}} & {\scriptstyle
-\f{819971}{19683}} {\scriptstyle -\f{19936}{243} \zetathree} & 0 \\
0 & 0 & \text{?} & \text{?} & \text{?} & \text{?} & {\scriptstyle
\f{1571264}{729}} & {\scriptstyle \f{3076372}{243}} & {\scriptstyle
-\f{16821944}{6561}} {\scriptstyle +\f{30464}{81} \zetathree} & 0 \\
0 & 0 & \text{?} & \text{?} & \text{?} & \text{?} & {\scriptstyle
-\f{1792768}{2187}} & {\scriptstyle -\f{3029846}{729}} & {\scriptstyle
-\f{17787368}{19683}} {\scriptstyle -\f{286720}{243} \zetathree} & 0
\\   
0 & 0 & 0 & 0 & 0 & 0 & \text{?} & 0 & 0 & 0 \\  
0 & 0 & 0 & 0 & 0 & 0 & \text{?} & \text{?} & 0 & 0 \\  
0 & 0 & 0 & 0 & 0 & 0 & 0 & 0 & {\scriptstyle -\f{9769}{27}} & 0 \\  
0 & 0 & 0 & 0 & 0 & 0 & 0 & 0 & 0 & {\scriptstyle -\f{9769}{27}}
\end{array}
\right ) \, .  
\eeq
The question marks stand for unknown entries that we have not
calculated. As far as the remaining entries are concerned, our results
for the one- and two-loop mixing of $Q_1$--$Q_6$ agree with those of
\cite{Chetyrkin:1998gb}, and therefore also with previous results
\cite{ADM4q} that were obtained in a different operator basis
\cite{differentbasis}. We also confirm other well-established results,
like the two-loop mixing of $Q_1$--$Q_6$ into $Q_7$--$Q_{10}$
\cite{twoloopmix}. See \cite{BMU} for the conversion to our basis. For
the  two-loop mixing of $Q_7$ and $Q_8$ we confirm the results of
\cite{misiak-munz}, and for the entries $\gamma^{(2)}_{7 i}$ and
$\gamma^{(2)}_{8 i}$, that is the three-loop mixing of four-quark
operators into magnetic ones, those of
\cite{Chetyrkin:1997vx}. Neither of the latter results have been    
checked before, with the exception of the self-mixing of $Q_7$  
\cite{broadhurst-grozin-gracey}. 

We recall that the ADM reported in \cite{Chetyrkin:1997vx} refers to
so-called effective coefficients. The relation with our results will
be discussed below. Finally, the entries $\gamma^{(2)}_{9 i}$ which
contain terms proportional to the Riemann zeta function $\zetathree$,
are entirely new. The ADM entries involving non-physical operators can
be readily obtained from the renormalization matrices provided in the
Appendix. Finally, let us mention here once again that for a NNLO
analysis of rare semileptonic $B$ decays one would also need to know
the self-mixing of $Q_1$--$Q_6$, which will be discussed in a separate
communication \cite{inpreparation}.

It might be useful to recall explicitly the relation between the ADM
in our basis and the ADM in an operator basis where $Q_7$--$Q_{10}$ are
not rescaled by $1/\gs^2$. The latter is frequently used for
phenomenological applications
\cite{gambino-misiak,misiak-buras,Chetyrkin:1997vx}. The Wilson
coefficients in that basis, $\widetilde{C}_i (\mu)$, are given by
\beq \label{eq:tildewilson}
\widetilde{C}_i (\mu) = 
\begin{cases}
C_i (\mu) \, , & \text{for $i = 1$--$6$} \, , \\[2mm]
\f{4 \pi}{\as} C_i (\mu) \, , & \text{for $i = 7$--$10$} \, , \
\end{cases}
\eeq 
while the coefficients in the expansion in powers of $\as$ of the
corresponding \ads \ $\widetilde{\gamma}_{ij}$, take the following
form:     
\beq \label{eq:tildegamma}
\widetilde{\gamma}^{(k - 1)}_{ij} = 
\begin{cases}
\gamma^{(k - 1)}_{ij} \, , & \text{for $i, j = 1$--$6$} \, , \\[2mm]  
\gamma^{(k)}_{ij} \, , & \text{for $i = 1$--$6 \, ,$ and $j =
7$--$10$} \, , \\[2mm]   
\gamma^{(k - 1)}_{ij} + 2 \beta_{k - 1} \delta_{i j} \, , & \text{for
$i, j = 7$--$10$} \, .  
\end{cases}
\eeq
Notice that the perturbative expansion of the \ads \
$\widetilde{\gamma}_{ij}$, that govern the evolution of 
$\widetilde{C}_i (\mu)$ already starts at zeroth order in $\as$, 
whereas $\gamma_{ij}$ contains no such terms. 

In the case of radiative $B$ decays it is sometimes convenient to use
effective Wilson coefficients $C_i^{\rm eff} (\mu)$ \cite{Buras:xp},
defined in such a way that the leading order $\btosgamma$ and
$\btosgluon$ matrix elements are proportional to $C_7^{\rm eff} (\mu)$
and $C_8^{\rm eff} (\mu)$, respectively. In particular, 
dropping the semileptonic operators which are irrelevant for radiative
$B$ decays, the effective Wilson coefficients in the operator basis of
\cite{Chetyrkin:1997vx} are given by       
\beq \label{eq:effectivewilson}
C^{\rm eff}_i (\mu) = 
\begin{cases}
C_i (\mu) \, , & \text{for $i = 1$--$6$} \, , \\[2mm] 
\f{4 \pi}{\as} C_i (\mu) + \sum^6_{j = 1} y^{(i)}_j C_j (\mu) \, , & 
\text{for $i = 7$--$8$} \, .  
\end{cases}
\eeq 
In the $\MSbar$ scheme with fully anticommuting $\gamma_5$ one has
$y^{(7)} = (0,0,-\f{1}{3}, -\f{4}{9}, -\f{20}{3}, -\f{80}{9})$ and
$y^{(8)} = (0, 0, 1, -\f{1}{6}, 20,-\f{10}{3})$. In consequence, the
first two coefficients in the perturbative expansion of the effective
ADM $\hat{\gamma}^{\rm eff}$, that governs the evolution of $C_i^{\rm
eff} (\mu)$ read 
\beq \label{eq:gammaeff1old}
\hat{\gamma}^{{\rm eff} (0)} = 
\left (
\begin{array}{cccccccccc}
-4 & {\scriptstyle \f{8}{3}} & 0 & {\scriptstyle -\f{2}{9}} & 0 & 0 &
{\scriptstyle -\f{208}{243}} & {\scriptstyle \f{173}{162}} \\  
12 & 0 & 0 & {\scriptstyle \f{4}{3}} & 0 & 0 & {\scriptstyle
\f{416}{81}} & {\scriptstyle \f{70}{27}} \\  
0 & 0 & 0 & {\scriptstyle -\f{52}{3}} & 0 & 2 & {\scriptstyle
-\f{176}{81}} & {\scriptstyle \f{14}{27}} \\  
0 & 0 & {\scriptstyle -\f{40}{9}} & {\scriptstyle -\f{100}{9}} &
{\scriptstyle \f{4}{9}} & {\scriptstyle \f{5}{6}} & {\scriptstyle
-\f{152}{243}} & {\scriptstyle -\f{587}{162}} \\  
0 & 0 & 0 & {\scriptstyle -\f{256}{3}} & 0 & 20 & {\scriptstyle
-\f{6272}{81}} & {\scriptstyle \f{6596}{27}} \\  
0 & 0 & {\scriptstyle -\f{256}{9}} & {\scriptstyle \f{56}{9}} &
{\scriptstyle \f{40}{9}} & {\scriptstyle -\f{2}{3}} & {\scriptstyle
\f{4624}{243}} & {\scriptstyle \f{4772}{81}} \\  
0 & 0 & 0 & 0 & 0 & 0 & {\scriptstyle \f{32}{3}} & 0 \\  
0 & 0 & 0 & 0 & 0 & 0 & {\scriptstyle -\f{32}{9}} & {\scriptstyle
\f{28}{3}}  
\end{array}
\right ) \, ,  
\eeq
and 
\beq \label{eq:gammaeff2old}
\hat{\gamma}^{{\rm eff} (1)} = 
\left (
\begin{array}{cccccccccc}
{\scriptstyle -\f{355}{9}} & {\scriptstyle -\f{502}{27}} &
{\scriptstyle -\f{1412}{243}} & {\scriptstyle -\f{1369}{243}} &
{\scriptstyle \f{134}{243}} & {\scriptstyle -\f{35}{162}} &
{\scriptstyle -\f{818}{243}} & {\scriptstyle \f{3779}{324}} \\  
{\scriptstyle -\f{35}{3}} & {\scriptstyle -\f{28}{3}} & {\scriptstyle
-\f{416}{81}} & {\scriptstyle \f{1280}{81}} & {\scriptstyle
\f{56}{81}} & {\scriptstyle \f{35}{27}} & {\scriptstyle \f{508}{81}} &
{\scriptstyle \f{1841}{108}} \\  
0 & 0 & {\scriptstyle -\f{4468}{81}} & {\scriptstyle -\f{31469}{81}} &
{\scriptstyle \f{400}{81}} & {\scriptstyle \f{3373}{108}} &
{\scriptstyle \f{22348}{243}} & {\scriptstyle \f{10178}{81}} \\  
0 & 0 & {\scriptstyle -\f{8158}{243}} & {\scriptstyle -\f{59399}{243}}
& {\scriptstyle \f{269}{486}} & {\scriptstyle \f{12899}{648}} &
{\scriptstyle -\f{17584}{243}} & {\scriptstyle -\f{172471}{648}} \\  
0 & 0 & {\scriptstyle -\f{251680}{81}} & {\scriptstyle
-\f{128648}{81}} & {\scriptstyle \f{23836}{81}} & {\scriptstyle
\f{6106}{27}} & {\scriptstyle \f{1183696}{729}} & {\scriptstyle
\f{2901296}{243}} \\  
0 & 0 & {\scriptstyle \f{58640}{243}} & {\scriptstyle -\f{26348}{243}}
& {\scriptstyle -\f{14324}{243}} & {\scriptstyle -\f{2551}{162}} &
{\scriptstyle \f{2480344}{2187}} & {\scriptstyle -\f{3296257}{729}} \\  
0 & 0 & 0 & 0 & 0 & 0 & {\scriptstyle \f{4688}{27}} & 0 \\  
0 & 0 & 0 & 0 & 0 & 0 & {\scriptstyle -\f{2192}{81}} & {\scriptstyle
\f{4063}{27}} 
\end{array}
\right ) \, , 
\eeq
which are in perfect agreement with \cite{Chetyrkin:1997vx}.

\section{Summary}

We have recalculated the complete $\ord (\as^2)$ mixing relevant for
the NLO analysis of radiative $B$ decays in the SM and some of its
extensions, confirming all previous results. In addition, we have also
calculated the $\ord (\as^2)$ three-loop mixing between four-quark and
semileptonic operators relevant for the NNLO analysis of rare
semileptonic $\btoslpluslminus$ transitions. This is a new result 
and provides one of the last missing pieces in the NNLO calculation
for $\BtoXslpluslminus$. We plan to complete the calculation of the
NNLO ADM required for this process and to study its phenomenological 
implications in a separate communication.  

The calculation involves the UV divergences of diagrams up to
three loops. Our results have been subject to several cross-checks,
following from: $i )$ the locality of the UV divergences,
$ii )$ the independence of the ADM from the external states used in
the calculation, $iii )$ the completeness of our operator basis, 
$iv)$ the gauge-parameter independence of the mixing among physical
operators, and $v )$ the absence of mixing of non-physical into
physical operators. We have also reproduced the full $\MSbar$
renormalization of QCD and QED up to the three-loop level.   

\subsubsection*{Acknowledgments}

We are grateful to M.~Misiak for his careful reading of the manuscript
and for many useful comments and discussions. Furthermore we would 
like to thank M.~Steinhauser for providing us with an updated version
of {\tt MATAD}. Finally we would like to thank A.~J.~Buras and
G.~Isidori for interesting discussions. The work of P.~G.\ is
supported by a Marie Curie Fellowship, contract
No. HPMF-CT-2000-01048. The work of M.~G.\ is supported by BMBF under
contract No. 05HT1WOA3. The work of U.~H.\ is supported by the U.S.
Department of Energy under contract No. DE-AC02-76CH03000. 

\section*{Appendix}

\subsection*{A.1  \ \ One- and Two-Loop QCD Renormalization
Constants}  

The renormalization of the conventional QCD Lagrangian containing a
massive bottom quark proceeds as usual. First, we introduce the
renormalized fields and variables via  
\beq \label{eq:qcdrenormalization}
\begin{split}
G^a_{\mu, B} & = Z^{1/2}_G G^a_\mu \, , \hspace{1cm} u^a_B = Z^{1/2}_u
u^a \, , \hspace{1cm} q_B  = Z^{1/2}_q q \, , \\ 
g_B & = Z_g g \, , \hspace{1.3cm} m_{b, B}  = Z_{\mb} \mb \, ,
\hspace{0.75cm} M_B  = Z_M M \, ,    
\end{split}
\eeq
where the subscript $B$ denotes bare quantities and $u^a$ are the
ghost fields. The gauge-parameter $\xi$ is kept unrenormalized. This
is legitimate, because the non-renormalization of the gauge-parameter
is guaranteed by the usual Slavnov-Taylor identity, which is
unaffected by the IR regularization  adopted for the Yang-Mills
theory. On the other hand, our IR rearrangement requires the
introduction of the gauge-variant subtraction in \Eq{eq:gluonmass},
which  can be interpreted as a counterterm for a fictitious gluon mass
$M$. 

Using the notation introduced in \Eq{eq:zfactors}, the $\MSbar$
renormalization constants at one-loop order take the following form  
\beq \label{eq:qcd11}
\begin{split}
Z^{(1, 1)}_G & = \left ( \f{13}{6} - \f{1}{2} \xi \right ) \ca -
\f{2}{3} \nf \, , \\ 
Z^{(1, 1)}_u & = \left ( \f{3}{4} - \f{1}{4} \xi \right ) \ca \, , \\ 
Z^{(1, 1)}_q & = - \xi \cf \, , \\ 
Z^{(1, 1)}_g & = -\f{11}{6} \ca + \f{1}{3} \nf \, , \\  
Z^{(1, 1)}_{\mb} & = -3 \cf \, , \\ 
Z^{(1, 1)}_M & = \left ( -\f{29}{24} - \f{1}{8} \xi \right ) \ca -
\f{2}{3} \nf \, ,   
\end{split}
\eeq
where $\ca = 3$ and $\cf = 4/3$ are the quadratic Casimir operators of
$\SUC$. As usual $\nf$ stands for the number of active quark
flavors. Our result for $Z^{(1, 1)}_M$ agrees with the expression 
given in \cite{Chetyrkin:1998fm}.  
  
At the two-loop level the poles of the $\MSbar$ renormalization
constants are given by  
\begin{align} \label{eq:qcd21}
Z^{(2, 1)}_G & = \left ( \f{59}{16} - \f{11}{16} \xi - \f{1}{8} \xi^2
\right ) \ca^2 - \cf \nf - \f{5}{4} \ca \nf \, , \non \\ 
Z^{(2, 1)}_u & = \left ( \f{95}{96} + \f{1}{32} \xi \right ) \ca^2 -
\f{5}{24} \ca \nf  \, , \non \\   
Z^{(2, 1)}_q & = \f{3}{4} \cf^2 - \left ( \f{25}{8} + \xi + \f{1}{8}
\xi^2 \right ) \cf \ca + \f{1}{2} \cf \nf \, , \non \\   
Z^{(2, 1)}_g & = -\f{17}{6} \ca^2 + \f{1}{2} \cf \nf + \f{5}{6} \ca
\nf \, , \non \\ 
Z^{(2, 1)}_{\mb} & = -\f{3}{4} \cf^2 - \f{97}{12} \cf \ca + \f{5}{6} \cf
\nf \, , \non \\  
Z^{(2, 1)}_M & = \left ( -\f{383}{192} - \f{7}{64} \xi - \f{3}{32}
\xi^2 \right ) \ca^2 + \left ( \f{1}{2} + \f{1}{4} \xi \right ) \cf
\nf + \left ( \f{5}{12} - \f{5}{16} \xi \right ) \ca \nf \, , 
\end{align}
and 
\beq \label{eq:qcd22}
\begin{split}
Z^{(2, 2)}_G & = \left ( -\f{13}{8} - \f{17}{24} \xi + \f{3}{16}
\xi^2 \right ) \ca^2 + \left ( \f{1}{2} + \f{1}{3} \xi \right ) \ca
\nf \, , \\ 
Z^{(2, 2)}_u & = \left ( -\f{35}{32} + \f{3}{32} \xi^2 \right ) \ca^2
+ \f{1}{4} \ca \nf \, , \\ 
Z^{(2, 2)}_q & = \f{1}{2} \xi^2 \cf^2 + \left ( \f{3}{4} \xi +
\f{1}{4} \xi^2 \right ) \cf \ca \, , \\ 
Z^{(2, 2)}_g & = \f{121}{24} \ca^2 - \f{11}{6} \ca \nf + 
\f{1}{6} \nf^2 \, , \\ 
Z^{(2, 2)}_{\mb} & = \f{9}{2} \cf^2 + \f{11}{2} \cf \ca - \cf \nf \, ,
\\
Z^{(2, 2)}_M  & = \left ( \f{1211}{384} + \f{59}{192} \xi + \f{5}{128}   
\xi^2 \right ) \ca^2 - \f{1}{2} \xi \cf \nf + \left ( \f{7}{12} -
\f{1}{24} \xi \right ) \ca \nf - \f{2}{3} \nf^2 \, . 
\end{split}
\eeq
Except for $Z^{(2, 1)}_M$ and $Z^{(2,2)}_M$, which have never been
given explicitly, our renormalization constants agree with the
results in the literature \cite{Muta}, if one bears in mind that 
the original papers contain some typing errors. We have also
calculated the three-loop renormalization constants
\cite{qcdqedrenormalization}, but we do not report them here, as they 
are not needed in our calculation. 
   
\subsection*{A.2  \ \ The Complete Operator Renormalization Matrix} 

The general structure of the operator renormalization matrix is
\beq \label{eq:znotation}
\hat{Z}^{(k, l)} = \left (
\begin{array}{ccc}
\hat{Z}^{(k, l)}_{\PP} & \hat{Z}^{(k, l)}_{\PN} & \hat{Z}^{(k,
l)}_{\PE} \\
\hat{Z}^{(k, l)}_{\NP} & \hat{Z}^{(k, l)}_{\NN} & \hat{Z}^{(k,
l)}_{\NE} \\ 
\hat{Z}^{(k, l)}_{\EP} & \hat{Z}^{(k, l)}_{\EN} & \hat{Z}^{(k,
l)}_{\EE} 
\end{array}
\right ) 
\, ,  
\eeq
where $P = 1$--$10$ denotes the physical operators, $N = 11$--$24$ the
EOM-vanishing operators, and $E = 25$--$32$ the evanescent operators.
Throughout this section we set $\nf = 5$.

The mixing of non-physical into physical operators must vanish at all
orders in $\as$. This is in fact  only a requirement on
the ADM, but we have seen in \Eq{eq:gammaexpansion} that the one-loop
renormalization  matrix $\hat{Z}^{(1, 1)}$ is proportional to
$\hat\gamma^{(0)}$, and therefore at one-loop it implies the vanishing
of $\hat{Z}^{(1, 1)}_{\NP}$ and $\hat{Z}^{(1, 1)}_{\EP}$. Since
$\hat\gamma^{(0)}$ for the physical operators can be found in
\Eq{eq:gamma0new}, it is sufficient to give here only  
the non-physical parts of $\hat Z^{(1, 0)}$ and $\hat{Z}^{(1, 1)}$. By 
definition, the only non-vanishing parts of $\hat{Z}^{(1, 0)}$  are
$\hat Z^{(1, 0)}_{\EP}$ and $\hat Z^{(1, 0)}_{\EN}$. We find  
{%
\renewcommand{\arraycolsep}{3.5pt}
\beq \label{eq:Z10EP}
\hat{Z}^{(1, 0)}_{\EP} =
\left (
\begin{array}{cccccccccc}
64 & {\scriptstyle \f{32}{3}} & 0 & {\scriptstyle \f{4}{9}} & 0 & 0 &
0 & 0 & {\scriptstyle \f{64}{27}} & 0 \\  
48 & -64 & 0 & {\scriptstyle -\f{8}{3}} & 0 & 0 & 0 & 0 &
{\scriptstyle \f{16}{9}} & 0 \\   
0 & 0 & {\scriptstyle \f{8960}{3}} & -2432 & {\scriptstyle
-\f{1280}{3}} & 320 & {\scriptstyle \f{64}{3}} & -64 & 16 & 0 \\  
0 & 0 & {\scriptstyle -\f{4480}{9}} & {\scriptstyle -\f{9464}{3}} &
{\scriptstyle \f{640}{9}} & {\scriptstyle \f{1280}{3}} & {\scriptstyle
\f{256}{9}} & {\scriptstyle \f{32}{3}} & {\scriptstyle -\f{256}{3}} &
0 \\  
3840 & 640 & 0 & 16 & 0 & 0 & 0 & 0 & {\scriptstyle \f{256}{3}} & 0 \\
2880 & -3840 & 0 & -96 & 0 & 0 & 0 & 0 & 64 & 0 \\  
0 & 0 & {\scriptstyle \f{609280}{3}} & -160768 & {\scriptstyle
-\f{98560}{3}} & 24640 & {\scriptstyle \f{512}{3}} & -512 & 544 & 0 \\
0 & 0 & {\scriptstyle -\f{304640}{9}} & {\scriptstyle -\f{630256}{3}}
& {\scriptstyle \f{49280}{9}} & {\scriptstyle \f{98560}{3}} &
{\scriptstyle \f{2048}{9}} & {\scriptstyle \f{256}{3}} & {\scriptstyle
-\f{11264}{3}} & 0 
\end{array}
\right ) \, ,  
\eeq
}%
and 
\beq \label{eq:Z10EN}
\hat{Z}^{(1, 0)}_{\EN} =
\left (
\begin{array}{cccccccccccccc}
{\scriptstyle \f{64}{27}} & {\scriptstyle -\f{4}{9}} & 0 & 0 & 0 & 0 &
0 & 0 & 0 & 0 & 0 & 0 & 0 & 0 \\  
{\scriptstyle \f{16}{9}} & {\scriptstyle \f{8}{3}} & 0 & 0 & 0 & 0 & 0
& 0 & 0 & 0 & 0 & 0 & 0 & 0 \\  
16 & 192 & 0 & 0 & 0 & 0 & 0 & 0 & 0 & 0 & 0 & 0 & 0 & 0 \\  
{\scriptstyle -\f{256}{3}} & 168 & 0 & 0 & 0 & 0 & 0 & 0 & 0 & 0 & 0 &
0 & 0 & 0 \\  
{\scriptstyle \f{256}{3}} & -16 & 0 & 0 & 0 & 0 & 0 & 0 & 0 & 0 & 0 &
0 & 0 & 0 \\  
64 & 96 & 0 & 0 & 0 & 0 & 0 & 0 & 0 & 0 & 0 & 0 & 0 & 0 \\  
544 & 8448 & 0 & 0 & 0 & 0 & 0 & 0 & 0 & 0 & 0 & 0 & 0 & 0 \\  
{\scriptstyle -\f{11264}{3}} & 6992 & 0 & 0 & 0 & 0 & 0 & 0 & 0 & 0 &
0 & 0 & 0 & 0 
\end{array}
\right ) \, .  
\eeq
The $6 \times 4$ block in the upper left corner of $\hat{Z}^{(1,
0)}_{\EP}$ agrees with the expression for the upper $6 \times 4$ block
of $\hat c$ given in Eq. (46) of \cite{Chetyrkin:1998gb}. 

The one-loop mixing of physical into non-physical operators is 
described by $\hat{Z}^{(1, 1)}_{\PN}$ and $\hat{Z}^{(1, 1)}_{\PE}$. We
get 
\beq \label{eq:Z11PN}
\hat{Z}^{(1, 1)}_{\PN} =
\left (
\begin{array}{cccccccccccccc}
{\scriptstyle -\f{16}{27}} & {\scriptstyle \f{1}{9}} & 0 & 0 & 0 & 0 &
0 & 0 & 0 & 0 & 0 & 0 & 0 & 0 \\  
{\scriptstyle -\f{4}{9}} & {\scriptstyle -\f{2}{3}} & 0 & 0 & 0 & 0 &
0 & 0 & 0 & 0 & 0 & 0 & 0 & 0 \\  
{\scriptstyle -\f{8}{9}} & {\scriptstyle -\f{4}{3}} & 0 & 0 & 0 & 0 &
0 & 0 & 0 & 0 & 0 & 0 & 0 & 0 \\  
{\scriptstyle \f{16}{27}} & {\scriptstyle -\f{28}{9}} & 0 & 0 & 0 & 0
& 0 & 0 & 0 & 0 & 0 & 0 & 0 & 0 \\  
{\scriptstyle -\f{56}{9}} & {\scriptstyle -\f{64}{3}} & 0 & 0 & 0 & 0
& 0 & 0 & 0 & 0 & 0 & 0 & 0 & 0 \\  
{\scriptstyle \f{256}{27}} & {\scriptstyle -\f{268}{9}} & 0 & 0 & 0 &
0 & 0 & 0 & 0 & 0 & 0 & 0 & 0 & 0 \\  
0 & 0 & 0 & 0 & 0 & 0 & 0 & 0 & 0 & 0 & 0 & 0 & 0 & 0 \\  
0 & 0 & -8 & 0 & 0 & 0 & {\scriptstyle -\f{9}{4}} & 0 & 0 & 0 & 0 & 0
& 0 & 0 \\  
0 & 0 & 0 & 0 & 0 & 0 & 0 & 0 & 0 & 0 & 0 & 0 & 0 & 0 \\  
0 & 0 & 0 & 0 & 0 & 0 & 0 & 0 & 0 & 0 & 0 & 0 & 0 & 0
\end{array}
\right ) \, ,   
\eeq
and 
\beq \label{eq:Z11PE}
\hat{Z}^{(1, 1)}_{\PE} =
\left (
\begin{array}{cccccccc}
{\scriptstyle \f{5}{12}} & {\scriptstyle \f{2}{9}} & 0 & 0 & 0 & 0 & 0
& 0 \\  
1 & 0 & 0 & 0 & 0 & 0 & 0 & 0 \\  
0 & 0 & 0 & 0 & 0 & 0 & 0 & 0 \\  
0 & 0 & 0 & 0 & 0 & 0 & 0 & 0 \\ 
0 & 0 & 0 & 1 & 0 & 0 & 0 & 0 \\ 
0 & 0 & {\scriptstyle \f{2}{9}} & {\scriptstyle \f{5}{12}} & 0 & 0 & 0
& 0 \\  
0 & 0 & 0 & 0 & 0 & 0 & 0 & 0 \\ 
0 & 0 & 0 & 0 & 0 & 0 & 0 & 0 \\  
0 & 0 & 0 & 0 & 0 & 0 & 0 & 0 \\ 
0 & 0 & 0 & 0 & 0 & 0 & 0 & 0
\end{array}
\right ) \, .  
\eeq
The $4 \times 6$ block in the upper left corner of $\hat{Z}^{(1,
1)}_{\PE}$ agrees with the expression for the $4 \times 6$ block in
the upper left corner of $\hat b$ given in Eq. (45) of
\cite{Chetyrkin:1998gb}.   

At one-loop we have moreover the mixing among EOM-vanishing operators,
given by    
{%
\renewcommand{\arraycolsep}{0.7pt}
\beq \label{eq:Z11NN}
\hat{Z}^{(1, 1)}_{\NN} =
\left (
\begin{array}{cccccccccccccc}
{\scriptstyle -\f{23}{3}} & 0 & 0 & 0 & 0 & 0 & 0 & 0 & 0 & 0 & 0 & 0 & 0 & 0 \\ 
 0 & {\scriptstyle -\f{9}{2}} & 0 & {\scriptstyle -\f{4}{3}} & 0 &
{\scriptstyle \f{3}{8}} & {\scriptstyle -\f{9}{8}} & {\scriptstyle
-\f{9}{8}} & {\scriptstyle -\f{3}{2}} & {\scriptstyle \f{3}{16}} &
{\scriptstyle \f{3}{2}} & 0 & {\scriptstyle \f{9}{4}} & {\scriptstyle
-\f{9}{16}} \\  
0 & 0 & {\scriptstyle -\f{11}{3}} {\scriptstyle -\f{4}{3} \xi} & 0 & 0
& 0 & 0 & 0 & 0 & 0 & 0 & 0 & 0 & 0 \\  
0 & 0 & {\scriptstyle 4} & {\scriptstyle -\f{23}{3}} {\scriptstyle
-\f{4}{3} \xi} & 0 & 0 & 0 & 0 & 0 & 0 & 0 & 0 & {\scriptstyle
\f{9}{4}} {\scriptstyle +\f{3}{4} \xi} & 0 \\  
0 & 0 & 0 & 0 & {\scriptstyle -\f{23}{3}} & 0 & 0 & 0 & 0 & 0 & 0 & 0
& 0 & 0 \\  
0 & 0 & {\scriptstyle -8} & {\scriptstyle 8} & {\scriptstyle
-\f{8}{9}} & {\scriptstyle -\f{17}{4}} & 0 & {\scriptstyle -\f{3}{4}}
& 0 & {\scriptstyle \f{1}{8}} & 0 & 0 & {\scriptstyle -\f{9}{2}} &
{\scriptstyle -\f{3}{8}} \\  
0 & 0 & {\scriptstyle \f{8}{3} \xi} & 0 & 0 & 0 & {\scriptstyle
-\f{11}{3}} & 0 & 0 & 0 & 0 & 0 & 0 & 0 \\  
0 & 0 & 0 & 0 & 0 & {\scriptstyle -\f{97}{24}} {\scriptstyle -\f{3}{8}
\xi} & 0 & {\scriptstyle -\f{31}{12}} & {\scriptstyle \f{13}{3}} &
{\scriptstyle -\f{61}{72}} {\scriptstyle -\f{1}{8} \xi} &
{\scriptstyle -\f{25}{3}} & {\scriptstyle -4} & {\scriptstyle
-\f{17}{2}} {\scriptstyle -\f{17}{6} \xi} & {\scriptstyle \f{95}{48}}
{\scriptstyle +\f{3}{16} \xi} \\  
0 & 0 & {\scriptstyle -4}{\scriptstyle -\f{8}{3} \xi} & {\scriptstyle
\f{8}{3} \xi} & 0 & {\scriptstyle \f{49}{24}} {\scriptstyle +\f{3}{8}
\xi} & 0 & {\scriptstyle -\f{13}{12}} & {\scriptstyle -8} &
{\scriptstyle \f{13}{72}} {\scriptstyle +\f{1}{8} \xi} & {\scriptstyle
\f{1}{3}} & 0 & {\scriptstyle \f{1}{2}} {\scriptstyle +\f{1}{6} \xi} &
{\scriptstyle \f{1}{48}} {\scriptstyle -\f{3}{16} \xi} \\  
0 & 0 & 0 & 0 & 0 & {\scriptstyle -\f{1}{2}} & 0 & 0 & {\scriptstyle
1} & {\scriptstyle -\f{65}{12}} {\scriptstyle -\f{3}{4} \xi} &
{\scriptstyle -4} & {\scriptstyle -3} & {\scriptstyle -3}{\scriptstyle
-\xi} & 0 \\  
0 & 0 & {\scriptstyle -\f{2}{3} \xi} & {\scriptstyle -2} {\scriptstyle
+\f{2}{3} \xi} & {\scriptstyle \f{2}{9}} & {\scriptstyle -\f{17}{24}}
& 0 & {\scriptstyle \f{1}{12}} & {\scriptstyle -\f{7}{24}} &
{\scriptstyle -\f{29}{144}} {\scriptstyle +\f{1}{8} \xi} &
{\scriptstyle -\f{13}{2}} & {\scriptstyle -\f{1}{4}} & {\scriptstyle
-\f{1}{4}} {\scriptstyle -\f{1}{12} \xi} & {\scriptstyle \f{29}{48}}
{\scriptstyle -\f{3}{16} \xi} \\  
0 & 0 & 0 & {\scriptstyle \f{4}{3} \xi} & 0 & 0 & 0 & 0 & 0 &
{\scriptstyle -\f{1}{8} \xi} & 0 & {\scriptstyle -\f{65}{12}} &
{\scriptstyle -\f{3}{4} \xi} & {\scriptstyle \f{3}{16} \xi} \\  
0 & 0 & 0 & 0 & 0 & 0 & 0 & 0 & 0 & 0 & 0 & 0 & {\scriptstyle
-\f{65}{12}} {\scriptstyle -\f{7}{12} \xi} & 0 \\  
0 & 0 & 0 & 0 & 0 & {\scriptstyle \f{5}{12}} & 0 & {\scriptstyle
-\f{5}{12}} & {\scriptstyle -\f{5}{6}} & {\scriptstyle \f{5}{72}} &
{\scriptstyle \f{10}{3}} & {\scriptstyle \f{5}{2}} & {\scriptstyle
\f{5}{2}} {\scriptstyle +\f{5}{6} \xi} & {\scriptstyle -\f{45}{8}}
{\scriptstyle -\f{3}{8} \xi} 
\end{array}
\right ) \, ,   
\eeq
}%
and the mixing among evanescent operators, which reads
\beq \label{eq:Z11EE}
\hat{Z}^{(1, 1)}_{\EE} =
\left (
\begin{array}{cccccccc}
-7 & {\scriptstyle -\f{4}{3}} & 0 & 0 & {\scriptstyle \f{5}{12}} &
{\scriptstyle \f{2}{9}} & 0 & 0 \\  
-6 & 0 & 0 & 0 & 1 & 0 & 0 & 0 \\  
0 & 0 & {\scriptstyle -\f{64}{3}} & -14 & 0 & 0 & 0 & 1 \\  
0 & 0 & {\scriptstyle -\f{28}{9}} & {\scriptstyle \f{13}{3}} & 0 & 0 &
{\scriptstyle \f{2}{9}} & {\scriptstyle \f{5}{12}} \\  
{  0} & {  0} & 0 & 0 & {\scriptstyle \f{13}{3}} &
{\scriptstyle -\f{28}{9}} & 0 & 0 \\   
{  0} & {  0} & 0 & 0 & -14 & {\scriptstyle
-\f{64}{3}} & 0 & 0 \\   
0 & 0 & {\scriptstyle \f{1792}{3}} & -784 & 0 & 0 & -64 & 38 \\  
0 & 0 & {\scriptstyle -\f{1568}{9}} & {\scriptstyle -\f{2212}{3}} & 0
& 0 & {\scriptstyle \f{76}{9}} & {\scriptstyle \f{166}{3}} 
\end{array}
\right ) \, .  
\eeq
The $4 \times 4$ block in the upper left corner of $\hat{Z}^{(1,
1)}_{\EE}$ agrees with the expression for the $4 \times 4$ block in
the upper left corner of $\hat d$ given in Eq. (47) of
\cite{Chetyrkin:1998gb}. The last block $\hat{Z}^{(1,1)}_{\NE}$, 
contains only zeros.  

Now we can proceed to the two-loop renormalization matrices. The 
non-vanishing blocks of $\hat{Z}^{(2,0)}$ are $\hat{Z}^{(2, 0)}_{\EP}$
and $\hat{Z}^{(2, 0)}_{\EN}$ for which we give only the rows
corresponding to the evanescent operators $Q_{25}$--$Q_{28}$. Our
results are     
\beq \label{eq:Z20EP}
\hat{Z}^{(2, 0)}_{25 \mbox{--} 28, P} = 
\left (
\begin{array}{cccccccccc}
{\scriptstyle \f{3908}{9}} & {\scriptstyle \f{2656}{27}} &
{\scriptstyle \f{7292}{243}} & {\scriptstyle \f{157}{243}} &
{\scriptstyle -\f{722}{243}} & {\scriptstyle \f{55}{81}} &
{\scriptstyle \f{1096}{243}} & {\scriptstyle -\f{761}{162}} &
{\scriptstyle \f{11392}{729}} & 0 \\  
{\scriptstyle \f{1760}{3}} & {\scriptstyle -\f{3584}{9}} &
{\scriptstyle \f{1616}{81}} & {\scriptstyle \f{3736}{81}} &
{\scriptstyle -\f{176}{81}} & {\scriptstyle -\f{110}{27}} &
{\scriptstyle -\f{2192}{81}} & {\scriptstyle -\f{454}{27}} &
{\scriptstyle -\f{4640}{243}} & 0 \\  
0 & 0 & {\scriptstyle \f{442528}{9}} & {\scriptstyle \f{589928}{27}}
& {\scriptstyle -\f{27680}{9}} & {\scriptstyle \f{15560}{9}} &
{\scriptstyle \f{9344}{9}} & {\scriptstyle -\f{18524}{3}} &
{\scriptstyle \f{23456}{81}} & 0 \\  
0 & 0 & {\scriptstyle -\f{145528}{9}} & {\scriptstyle \f{860864}{81}}
& {\scriptstyle \f{41276}{27}} & {\scriptstyle \f{108475}{54}} &
{\scriptstyle -\f{2416}{3}} & {\scriptstyle -\f{177}{2}} &
{\scriptstyle -\f{152512}{243}} & 0 
\end{array}
\right ) \, ,  
\eeq
and
\beq \label{eq:Z20EN}
\hat{Z}^{(2, 0)}_{25 \mbox{--} 28, N} =
\left (
\begin{array}{cccccccccccccc}
{\scriptstyle \f{11392}{729}} & {\scriptstyle -\f{3101}{486}} &
{\scriptstyle -\f{4}{9}} & {\scriptstyle -\f{4}{27}} & {\scriptstyle
-\f{376}{243}} & \text{?} & \text{?} & \text{?} & \text{?} & \text{?}
& \text{?} & \text{?} & \text{?} & \text{?} \\  
{\scriptstyle -\f{4640}{243}} & {\scriptstyle -\f{949}{81}} &
{\scriptstyle \f{8}{3}} & {\scriptstyle \f{8}{9}} & {\scriptstyle
\f{752}{81}} & \text{?} & \text{?} & \text{?} & \text{?} & \text{?} &
\text{?} & \text{?} & \text{?} & \text{?} \\  
{\scriptstyle \f{23456}{81}} & {\scriptstyle -\f{35120}{27}} &
{\scriptstyle \f{3712}{3}} & {\scriptstyle -\f{64}{9}} & {\scriptstyle
-\f{1088}{9}} & \text{?} & \text{?} & \text{?} & \text{?} & \text{?} &
\text{?} & \text{?} & \text{?} & \text{?} \\  
{\scriptstyle -\f{152512}{243}} & {\scriptstyle \f{337192}{81}} &
{\scriptstyle -\f{56}{9}} & {\scriptstyle -\f{1768}{27}} &
{\scriptstyle \f{688}{9}} & \text{?} & \text{?} & \text{?} & \text{?}
& \text{?} & \text{?} & \text{?} & \text{?} & \text{?} 
\end{array}
\right ) \, ,  
\eeq
where the question marks correspond to entries that we have not
calculated. Notice that only the $4 \times 4$ block to the right of 
$\hat{Z}^{(2, 0)}_{25 \mbox{--} 28, P}$ is needed to determine the  
$\ord (\as^3)$ mixing of $Q_1$--$Q_6$ into $Q_7$--$Q_{10}$. On the
other hand the $6 \times 4$ block to the left is necessary to find the
three-loop self-mixing of $Q_1$--$Q_6$ which we shall present
elsewhere \cite{inpreparation}. Finally the mixing of evanescent 
operators into EOM-vanishing ones, described by $\hat{Z}^{(2, 0)}_{25 
\mbox{--} 28, N}$, is not necessary, but given for completeness here.

Since $\hat{Z}^{(2, 2)}$ is completely determined by the one-loop
mixing, we give only the non-vanishing building blocks of
$\hat{Z}^{(2, 1)}$, namely 
\beq \label{eq:Z21PP}
\hat{Z}^{(2, 1)}_{\PP} =
\left (
\begin{array}{cccccccccc}
{\scriptstyle \f{317}{36}} & {\scriptstyle -\f{515}{54}} &
{\scriptstyle -\f{353}{243}} & {\scriptstyle -\f{1567}{972}} &
{\scriptstyle \f{67}{486}} & {\scriptstyle -\f{35}{648}} &
{\scriptstyle -\f{58}{243}} & {\scriptstyle \f{167}{648}} &
{\scriptstyle -\f{64}{729}} & 0 \\  
{\scriptstyle \f{349}{12}} & 3 & {\scriptstyle -\f{104}{81}} &
{\scriptstyle \f{338}{81}} & {\scriptstyle \f{14}{81}} & {\scriptstyle
\f{35}{108}} & {\scriptstyle \f{116}{81}} & {\scriptstyle \f{19}{27}}
& {\scriptstyle \f{776}{243}} & 0 \\  
0 & 0 & {\scriptstyle -\f{1117}{81}} & {\scriptstyle -\f{31469}{324}}
& {\scriptstyle \f{100}{81}} & {\scriptstyle \f{3373}{432}} &
{\scriptstyle \f{16}{81}} & {\scriptstyle \f{92}{27}} & {\scriptstyle
-\f{1688}{243}} & 0 \\  
0 & 0 & {\scriptstyle -\f{4079}{486}} & {\scriptstyle -\f{59399}{972}}
& {\scriptstyle \f{269}{1944}} & {\scriptstyle \f{12899}{2592}} &
{\scriptstyle -\f{50}{243}} & {\scriptstyle -\f{1409}{648}} &
{\scriptstyle -\f{548}{729}} & 0 \\  
0 & 0 & {\scriptstyle -\f{83080}{81}} & {\scriptstyle -\f{159926}{81}}
& {\scriptstyle \f{8839}{81}} & {\scriptstyle \f{14573}{54}} &
{\scriptstyle -\f{464}{81}} & {\scriptstyle \f{3407}{27}} &
{\scriptstyle -\f{31376}{243}} & 0 \\  
0 & 0 & {\scriptstyle \f{70100}{243}} & {\scriptstyle
-\f{231956}{243}} & {\scriptstyle -\f{11501}{243}} & {\scriptstyle
\f{78089}{648}} & {\scriptstyle -\f{836}{243}} & {\scriptstyle
-\f{1081}{81}} & {\scriptstyle -\f{21128}{729}} & 0 \\  
0 & 0 & 0 & 0 & 0 & 0 & {\scriptstyle \f{650}{27}} & 0 & 0 & 0 \\  
0 & 0 & 0 & 0 & 0 & 0 & {\scriptstyle -\f{548}{81}} & {\scriptstyle
\f{1975}{108}} & 0 & 0 \\  
0 & 0 & 0 & 0 & 0 & 0 & 0 & 0 & {\scriptstyle -\f{58}{3}} & 0 \\  
0 & 0 & 0 & 0 & 0 & 0 & 0 & 0 & 0 & {\scriptstyle -\f{58}{3}} 
\end{array}
\right ) \, ,  
\eeq
{%
\renewcommand{\arraycolsep}{1.25pt}
\beq \label{eq:Z21PN}
\hat{Z}^{(2, 1)}_{\PN} =
\left (
\begin{array}{cccccccccccccc}
{\scriptstyle -\f{64}{729}} & {\scriptstyle \f{3671}{1944}} &
{\scriptstyle \f{1}{9}} & {\scriptstyle -\f{1}{27}} & {\scriptstyle
\f{22}{243}} & {\scriptstyle -\f{65}{2592}} & {\scriptstyle
-\f{1}{96}} & {\scriptstyle \f{1}{32}} & {\scriptstyle \f{1}{72}} &
{\scriptstyle -\f{1}{192}} & {\scriptstyle -\f{1}{72}} & 0 &
{\scriptstyle \f{1}{48}} & {\scriptstyle \f{1}{64}} \\  
{\scriptstyle \f{776}{243}} & {\scriptstyle -\f{1889}{324}} &
{\scriptstyle -\f{2}{3}} & {\scriptstyle \f{2}{9}} & {\scriptstyle
-\f{44}{81}} & {\scriptstyle -\f{259}{432}} & {\scriptstyle \f{1}{16}}
& {\scriptstyle -\f{3}{16}} & {\scriptstyle -\f{1}{12}} &
{\scriptstyle \f{1}{32}} & {\scriptstyle \f{1}{12}} & 0 &
{\scriptstyle -\f{1}{8}} & {\scriptstyle -\f{3}{32}} \\  
{\scriptstyle -\f{1688}{243}} & {\scriptstyle \f{1351}{162}} &
{\scriptstyle -\f{16}{3}} & {\scriptstyle \f{4}{9}} & {\scriptstyle
\f{20}{81}} & {\scriptstyle -\f{259}{216}} & -1 & {\scriptstyle
-\f{3}{8}} & {\scriptstyle -\f{1}{6}} & {\scriptstyle \f{1}{16}} &
{\scriptstyle \f{1}{6}} & 0 & {\scriptstyle -\f{1}{4}} & {\scriptstyle
-\f{3}{16}} \\  
{\scriptstyle -\f{548}{729}} & {\scriptstyle -\f{3559}{972}} &
{\scriptstyle -\f{22}{9}} & {\scriptstyle -\f{32}{27}} & {\scriptstyle
-\f{130}{243}} & {\scriptstyle -\f{625}{648}} & {\scriptstyle
-\f{67}{48}} & {\scriptstyle -\f{11}{4}} & {\scriptstyle -\f{26}{9}} &
{\scriptstyle \f{11}{24}} & {\scriptstyle \f{26}{9}} & 0 &
{\scriptstyle \f{19}{6}} & {\scriptstyle -\f{11}{8}} \\  
{\scriptstyle -\f{31376}{243}} & {\scriptstyle \f{2054}{81}} &
{\scriptstyle -\f{304}{3}} & {\scriptstyle \f{32}{3}} & {\scriptstyle
\f{320}{81}} & {\scriptstyle -\f{950}{27}} & {\scriptstyle -\f{35}{2}}
& -3 & {\scriptstyle \f{4}{3}} & {\scriptstyle \f{1}{2}} &
{\scriptstyle -\f{4}{3}} & 0 & -10 & {\scriptstyle -\f{3}{2}} \\  
{\scriptstyle -\f{21128}{729}} & {\scriptstyle -\f{32669}{486}} &
{\scriptstyle -\f{148}{9}} & {\scriptstyle \f{4}{9}} & {\scriptstyle
-\f{1684}{243}} & {\scriptstyle -\f{12817}{648}} & {\scriptstyle
-\f{35}{24}} & {\scriptstyle -\f{131}{8}} & {\scriptstyle
-\f{259}{18}} & {\scriptstyle \f{131}{48}} & {\scriptstyle
\f{259}{18}} & 0 & {\scriptstyle \f{125}{12}} & {\scriptstyle
-\f{131}{16}} \\  
0 & 0 & 0 & 0 & 0 & 0 & 0 & 0 & 0 & 0 & 0 & 0 & 0 & 0 \\  
0 & 0 & {\scriptstyle -\f{995}{18}} {\scriptstyle -\f{7}{9} \xi} & 0 &
0 & 0 & {\scriptstyle -\f{43}{8}} {\scriptstyle -\f{27}{32} \xi} & 0 &
0 & 0 & 0 & 0 & 0 & 0 \\   
0 & 0 & 0 & 0 & 0 & 0 & 0 & 0 & 0 & 0 & 0 & 0 & 0 & 0 \\  
0 & 0 & 0 & 0 & 0 & 0 & 0 & 0 & 0 & 0 & 0 & 0 & 0 & 0
\end{array}
\right ) \, ,  
\eeq
}%
and
\beq \label{eq:Z21PE}
\hat{Z}^{(2, 1)}_{\PE} =
\left (
\begin{array}{cccccccc}
{\scriptstyle \f{4493}{864}} & {\scriptstyle -\f{49}{648}} & 0 & 0 &
{\scriptstyle \f{1}{384}} & {\scriptstyle -\f{35}{864}} & 0 & 0 \\  
{\scriptstyle \f{1031}{144}} & {\scriptstyle \f{8}{9}} & 0 & 0 &
{\scriptstyle -\f{35}{192}} & {\scriptstyle -\f{7}{72}} & 0 & 0 \\  
0 & 0 & {\scriptstyle -\f{7}{72}} & {\scriptstyle -\f{35}{192}} & 0 &
0 & 0 & 0 \\  
0 & 0 & {\scriptstyle -\f{35}{864}} & {\scriptstyle \f{1}{384}} & 0 &
0 & 0 & 0 \\  
0 & 0 & {\scriptstyle \f{23}{18}} & {\scriptstyle \f{449}{36}} & 0 & 0
& {\scriptstyle -\f{7}{72}} & {\scriptstyle -\f{35}{192}} \\  
0 & 0 & {\scriptstyle \f{179}{162}} & {\scriptstyle \f{463}{108}} & 0
& 0 & {\scriptstyle -\f{35}{864}} & {\scriptstyle \f{1}{384}} \\  
0 & 0 & 0 & 0 & 0 & 0 & 0 & 0 \\ 
0 & 0 & 0 & 0 & 0 & 0 & 0 & 0 \\ 
0 & 0 & 0 & 0 & 0 & 0 & 0 & 0 \\ 
0 & 0 & 0 & 0 & 0 & 0 & 0 & 0
\end{array}
\right ) \, . 
\eeq
Similarly to what happens in the case of $\hat{Z}^{(2, 0)}_{\PE}$ not
all entries of $\hat{Z}^{(2, 1)}_{\PE}$ are needed to find the
$\ord (\as^3)$ ADM of physical operators considered in this
article. Needless to say, the mixing of physical into EOM-vanishing
operators, described by $\hat{Z}^{(2, 1)}_{\PN}$, is not required to
determine the mixing of physical operators at the three-loop
level. However, some entries are important to verify the $\ord
(\as^2)$ mixing of magnetic into non-physical operators which has been
discussed in part in \cite{misiak-munz}.    

As far as the mixing among EOM-vanishing operators is concerned, we
have calculated only the first two rows of the corresponding matrix
$\hat{Z}^{(2, 1)}_{\NN}$. We find   
\beq \label{eq:Z21NN}
\hat{Z}^{(2, 1)}_{11 \mbox{--} 12, N} = 
\left (
\begin{array}{cccccccccccccc}
{\scriptstyle -\f{58}{3}} & 0 & 0 & 0 & 0 & \text{?} & \text{?} &
\text{?} & \text{?} & \text{?} & \text{?} & \text{?} & \text{?} &
\text{?} \\  
0 & {\scriptstyle -\f{149}{16}} & {\scriptstyle \f{13}{36}}
{\scriptstyle -\f{7}{36} \xi} & {\scriptstyle -\f{11}{2}}
{\scriptstyle -\f{7}{36} \xi} & {\scriptstyle -\f{1}{6}} & \text{?} &
\text{?} & \text{?} & \text{?} & \text{?} & \text{?} & \text{?} &
\text{?} & \text{?} 
\end{array}
\right ) \, ,   
\eeq
where the question marks stand for entries that we have not
calculated. 

In the case of the mixing of evanescent into other operators, we have
calculated only the first four rows of the corresponding matrices
$\hat{Z}^{(2, 1)}_{\EP}$, $\hat{Z}^{(2, 1)}_{\EN}$ and $\hat{Z}^{(2,
1)}_{\EE}$. We get  
\beq \label{eq:Z21EP}
\hat{Z}^{(2, 1)}_{25 \mbox{--} 28, P} =
\left (
\begin{array}{cccccccccc}
{\scriptstyle \f{1760}{3}} & {\scriptstyle -\f{2576}{9}} &
{\scriptstyle -\f{40}{81}} & {\scriptstyle -\f{814}{81}} &
{\scriptstyle \f{4}{81}} & {\scriptstyle \f{5}{54}} & 0 & 0 &
{\scriptstyle -\f{5824}{243}} & 0 \\  
1304 & {\scriptstyle \f{1696}{3}} & {\scriptstyle \f{80}{27}} &
{\scriptstyle \f{8}{27}} & {\scriptstyle -\f{8}{27}} & {\scriptstyle
-\f{5}{9}} & 0 & 0 & {\scriptstyle \f{1712}{81}} & 0 \\  
0 & 0 & -56320 & {\scriptstyle -\f{132848}{3}} & 8512 & 7600 &
{\scriptstyle -\f{1088}{3}} & 992 & {\scriptstyle -\f{6992}{9}} & 0 \\
0 & 0 & {\scriptstyle \f{109520}{9}} & {\scriptstyle -\f{127570}{3}}
& {\scriptstyle -\f{16568}{9}} & {\scriptstyle \f{23255}{3}} &
{\scriptstyle -\f{512}{9}} & {\scriptstyle \f{80}{3}} & {\scriptstyle
\f{2432}{3}} & 0 
\end{array}
\right ) \, ,  
\eeq
\beq \label{eq:Z21EN}
\hat{Z}^{(2, 1)}_{25 \mbox{--} 28, N} =
\left (
\begin{array}{cccccccccccccc}
{\scriptstyle -\f{5824}{243}} & {\scriptstyle \f{739}{81}} & 0 &
{\scriptstyle \f{8}{27}} & 0 & \text{?} & \text{?} & \text{?} &
\text{?} & \text{?} & \text{?} & \text{?} & \text{?} & \text{?} \\  
{\scriptstyle \f{1712}{81}} & {\scriptstyle \f{142}{27}} & 0 &
{\scriptstyle -\f{16}{9}} & 0 & \text{?} & \text{?} & \text{?} &
\text{?} & \text{?} & \text{?} & \text{?} & \text{?} & \text{?} \\  
{\scriptstyle -\f{6992}{9}} & {\scriptstyle \f{2048}{3}} & 256 & -128
& 0 & \text{?} & \text{?} & \text{?} & \text{?} & \text{?} & \text{?}
& \text{?} & \text{?} & \text{?} \\  
{\scriptstyle \f{2432}{3}} & {\scriptstyle -\f{1300}{3}} &
{\scriptstyle -\f{128}{3}} & -112 & 0 & \text{?} & \text{?} & \text{?}
& \text{?} & \text{?} & \text{?} & \text{?} & \text{?} & \text{?} 
\end{array}
\right ) \, ,   
\eeq
and 
\beq \label{eq:Z21EE}
\hat{Z}^{(2, 1)}_{25 \mbox{--} 28, E} =
\left (
\begin{array}{cccccccccccccc}
{\scriptstyle \f{1615}{24}} & {\scriptstyle -\f{1021}{27}} & 0 & 0 &
{\scriptstyle \f{917}{216}} & {\scriptstyle \f{142}{81}} & 0 & 0 \\  
{\scriptstyle \f{599}{6}} & {\scriptstyle \f{715}{9}} & 0 & 0 &
{\scriptstyle \f{277}{18}} & {\scriptstyle \f{17}{6}} & 0 & 0 \\  
0 & 0 & {\scriptstyle \f{5263}{27}} & {\scriptstyle \f{2255}{36}} & 0
& 0 & {\scriptstyle -\f{13}{9}} & {\scriptstyle \f{3041}{144}} \\  
0 & 0 & {\scriptstyle -\f{10489}{162}} & {\scriptstyle \f{57317}{108}}
& 0 & 0 & {\scriptstyle \f{1961}{648}} & {\scriptstyle \f{1427}{864}}
\end{array}
\right ) \, .  
\eeq
Here once again question marks denote entries that we have not
computed. Clearly, the mixing of evanescent into other operators does
not affect the $\ord (\as^3)$ mixing of physical operators at all, and
thus is given here only for completeness.    

At the three-loop level we have calculated only a small subset of
entries of $\hat{Z}^{(3, 1)}$ which are summarized below. Again
$\hat{Z}^{(3, 2)}$ and $\hat{Z}^{(3, 3)}$ can in principle be obtained
using \Eqs{eq:localitychecks}. The single poles we have calculated
read 
\beq \label{eq:Z31PP}
\hat{Z}^{(3, 1)}_{\PP} =
\left (
\begin{array}{cccccccccc}
\text{?} & \text{?} & \text{?} & \text{?} & \text{?} & \text{?} &
{\scriptstyle -\f{15659}{6561}} & {\scriptstyle -\f{9625}{8748}} &
{\scriptstyle -\f{248315}{59049}} {\scriptstyle +\f{3488}{729}
\zetathree} & 0 \\  
\text{?} & \text{?} & \text{?} & \text{?} & \text{?} & \text{?} &
{\scriptstyle \f{13390}{2187}} & {\scriptstyle \f{5749}{5832}} &
{\scriptstyle -\f{54656}{19683}} {\scriptstyle -\f{1792}{243}
\zetathree} & 0 \\  
0 & 0 & \text{?} & \text{?} & \text{?} & \text{?} & {\scriptstyle
\f{35528}{2187}} & {\scriptstyle \f{35113}{729}} & {\scriptstyle
-\f{461338}{19683}} {\scriptstyle +\f{1600}{243} \zetathree} & 0 \\   
0 & 0 & \text{?} & \text{?} & \text{?} & \text{?} & {\scriptstyle
-\f{95551}{6561}} & {\scriptstyle -\f{1356773}{34992}} & {\scriptstyle
-\f{888497}{118098}} {\scriptstyle -\f{9968}{729} \zetathree} & 0 \\
0 & 0 & \text{?} & \text{?} & \text{?} & \text{?} & {\scriptstyle
\f{670864}{2187}} & {\scriptstyle \f{3116449}{1458}} & {\scriptstyle
-\f{17938948}{19683}} {\scriptstyle +\f{15232}{243} \zetathree} & 0 \\  
0 & 0 & \text{?} & \text{?} & \text{?} & \text{?} & {\scriptstyle
-\f{516836}{6561}} & {\scriptstyle -\f{20383751}{17496}} &
{\scriptstyle -\f{27731962}{59049}} {\scriptstyle -\f{143360}{729} 
\zetathree} & 0 \\  
0 & 0 & 0 & 0 & 0 & 0 & \text{?} & 0 & 0 & 0 \\  
0 & 0 & 0 & 0 & 0 & 0 & \text{?} & \text{?} & 0 & 0 \\  
0 & 0 & 0 & 0 & 0 & 0 & 0 & 0 & {\scriptstyle -\f{9769}{162}} & 0 \\  
0 & 0 & 0 & 0 & 0 & 0 & 0 & 0 & 0 & {\scriptstyle -\f{9769}{162}}
\end{array}
\right ) \, ,  
\eeq
and
{%
\renewcommand{\arraycolsep}{0.5pt}
\beq \label{eq:Z31PN}
\hat{Z}^{(3, 1)}_{P, 11 \mbox{--} 15} =
\left (
\begin{array}{ccccc}
{\scriptstyle -\f{248315}{59049}} {\scriptstyle +\f{3488}{729}
\zetathree} & {\scriptstyle \f{2737385}{314928} + \f{1}{16} \xi
-\f{437}{243} \zetathree} & {\scriptstyle \f{1853}{324}} {\scriptstyle
+\f{287}{1944} \xi} & {\scriptstyle \f{1955}{13122}} {\scriptstyle
-\f{1}{486} \xi} {\scriptstyle +\f{80}{81} \zetathree} & {\scriptstyle
\f{31883}{26244}} {\scriptstyle +\f{88}{243} \zetathree} \\  
{\scriptstyle -\f{54656}{19683}} {\scriptstyle -\f{1792}{243}
\zetathree} & {\scriptstyle -\f{529049}{13122} -\f{3}{8} \xi
+\f{2818}{81} \zetathree} & {\scriptstyle -\f{521}{54}} {\scriptstyle 
-\f{287}{324} \xi} & {\scriptstyle \f{23722}{2187}} {\scriptstyle
+\f{1}{81} \xi} {\scriptstyle -\f{160}{27} \zetathree} & {\scriptstyle
-\f{15305}{4374}} {\scriptstyle +\f{112}{81} \zetathree} \\  
{\scriptstyle -\f{461338}{19683}} {\scriptstyle +\f{1600}{243}
\zetathree} & {\scriptstyle \f{5097593}{52488} -\f{3}{4} \xi
+\f{8336}{81} \zetathree} & {\scriptstyle -\f{247}{9}} {\scriptstyle 
-\f{403}{81} \xi} & {\scriptstyle \f{62024}{2187}} {\scriptstyle
+\f{2}{81} \xi} {\scriptstyle -\f{320}{27} \zetathree} & {\scriptstyle
\f{3271}{2187}} {\scriptstyle -\f{16}{81} \zetathree} \\  
{\scriptstyle -\f{888497}{118098}} {\scriptstyle -\f{9968}{729}
\zetathree} & {\scriptstyle -\f{4401395}{314928} -\f{7}{4} \xi 
+\f{97237}{486} \zetathree} & {\scriptstyle -\f{8389}{162}}
{\scriptstyle -\f{5899}{972} \xi} & {\scriptstyle \f{378025}{13122}}
{\scriptstyle +\f{523}{486} \xi} {\scriptstyle -\f{2240}{81}
\zetathree} & {\scriptstyle -\f{48901}{13122}} {\scriptstyle
+\f{32}{243} \zetathree} \\  
{\scriptstyle -\f{17938948}{19683}} {\scriptstyle +\f{15232}{243}
\zetathree} & {\scriptstyle \f{50765453}{13122} - 12 \, \xi 
+\f{139856}{81} \zetathree} & {\scriptstyle -\f{61396}{81}}
{\scriptstyle -\f{6410}{81} \xi} & {\scriptstyle \f{1083404}{2187}} 
{\scriptstyle -\f{100}{81} \xi} {\scriptstyle -\f{5120}{27}
\zetathree} & {\scriptstyle \f{99716}{2187}} {\scriptstyle
-\f{832}{81} \zetathree} \\  
{\scriptstyle -\f{27731962}{59049}} {\scriptstyle -\f{143360}{729}
\zetathree} & {\scriptstyle \f{9909208}{19683} -\f{67}{4} \xi
+\f{418166}{243} \zetathree} & {\scriptstyle -\f{34099}{243}}
{\scriptstyle -\f{19915}{486} \xi} & {\scriptstyle \f{2372774}{6561}}
{\scriptstyle +\f{1190}{243} \xi} {\scriptstyle -\f{21440}{81}
\zetathree} & {\scriptstyle -\f{428893}{6561}} {\scriptstyle
-\f{208}{243} \zetathree} \\  
0 & 0 & 0 & 0 & 0  \\ 
0 & 0 & \text{?} & 0 & 0 \\
0 & 0 & 0 & 0 & 0 \\  
0 & 0 & 0 & 0 & 0 
\end{array}
\right ) \, .  
\eeq
}%

\def\pl#1#2#3{{Phys. Lett. }{\bf B#1~}(19#2)~#3}
\def\zp#1#2#3{{Z. Phys. }{\bf C#1~}(19#2)~#3}
\def\prl#1#2#3{{Phys. Rev. Lett. }{\bf #1~}(19#2)~#3}
\def\rmp#1#2#3{{Rev. Mod. Phys. }{\bf #1~}(19#2)~#3}
\def\prep#1#2#3{{Phys. Rep. }{\bf #1~}(19#2)~#3}
\def\pr#1#2#3{{Phys. Rev. }{\bf D#1~}(19#2)~#3}
\def\np#1#2#3{{Nucl. Phys. }{\bf B#1~}(19#2)~#3}
\def\mpl#1#2#3{{Mod. Phys. Lett. }{\bf #1~}(19#2)~#3}

\end{document}